\documentclass{raa}          

\usepackage{graphicx,times,epsf}             
\usepackage{amsmath}
\usepackage{makecell}
\usepackage{multirow}

\begin{document}
   \title{$^{12}$CO emission from the Red Rectangle}
   \volnopage{Vol.0 (200x) No.0, 000--000}      
   \setcounter{page}{1}          
   \author{Pham Tuan Anh\inst{}
          \and
          Pham Ngoc Diep\inst{}
          \and
          Do Thi Hoai\inst{}
          \and
          Pham Tuyet Nhung\inst{}
          \and
          Nguyen Thi Phuong\inst{}
          \and
          Nguyen Thi Thao\inst{}
          \and
          Pierre Darriulat\inst{}
     }     
   \institute{Department of Astrophysics, Vietnam National Satellite Center, VAST, 18 Hoang Quoc Viet, Cau Giay, Hanoi, Vietnam
             }
   \date{}
   \abstract{Observations of an unprecedented quality made by ALMA on the Red Rectangle of \mbox{CO(3-2)} and \mbox{CO(6-5)} emissions are analysed jointly with the aim of obtaining as simple as possible a description of the gas morphology and kinematics. Evidence is found for polar conical outflows and for a broad equatorial torus in rotation and expansion. Simple models of both are proposed. Comparing \mbox{CO(6-5)} and \mbox{CO(3-2)} emissions provides evidence for a strong temperature enhancement over the polar outflows. Continuum emission (dust) is seen to be enhanced in the equatorial region. Observed asymmetries are briefly discussed.
   \keywords{stars: AGB and post-AGB $-$ circumstellar matter $-$ radio-lines: stars: $-$ planetary nebulae: individual (Red Rectangle)}
   }
   \authorrunning{P. T. Anh et al.}            
   \titlerunning{}  
   \maketitle

%

\section{Introduction}
\begin{figure*}[htb]
\begin{center}
\begin{tabular}{p{5.1cm} p{4.1cm} p{4.1cm}}
\vspace{0pt}\includegraphics[scale=.195,trim=0. -1.5cm. .5cm -1.2cm,clip]{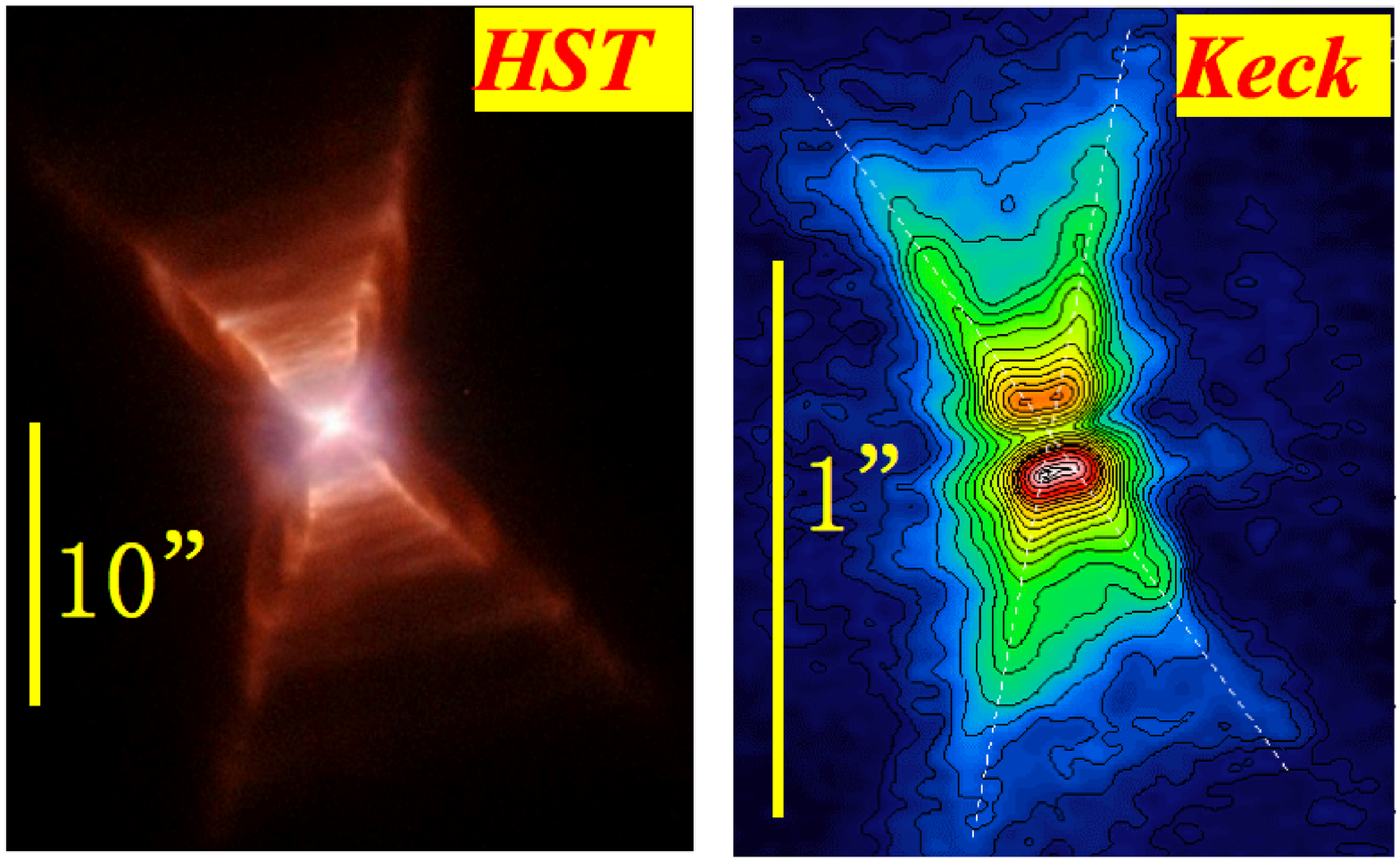} &
\vspace{-.15cm}\includegraphics[scale=.26,trim=1.0cm 0. 2.cm 0.cm, clip]{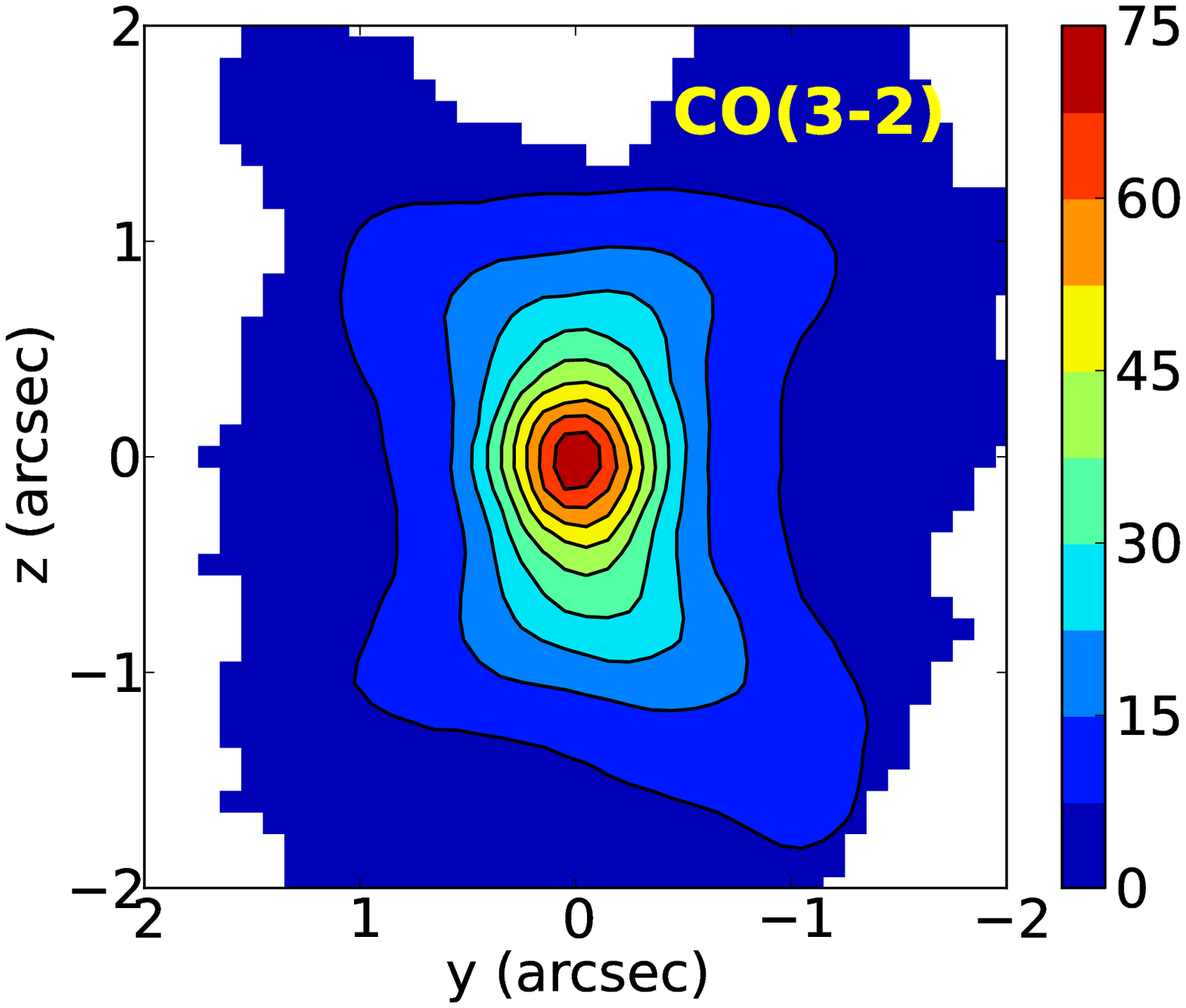} &
\vspace{-.15cm}\includegraphics[scale=.26,trim=0.5cm 0. 2.cm 0.cm, clip]{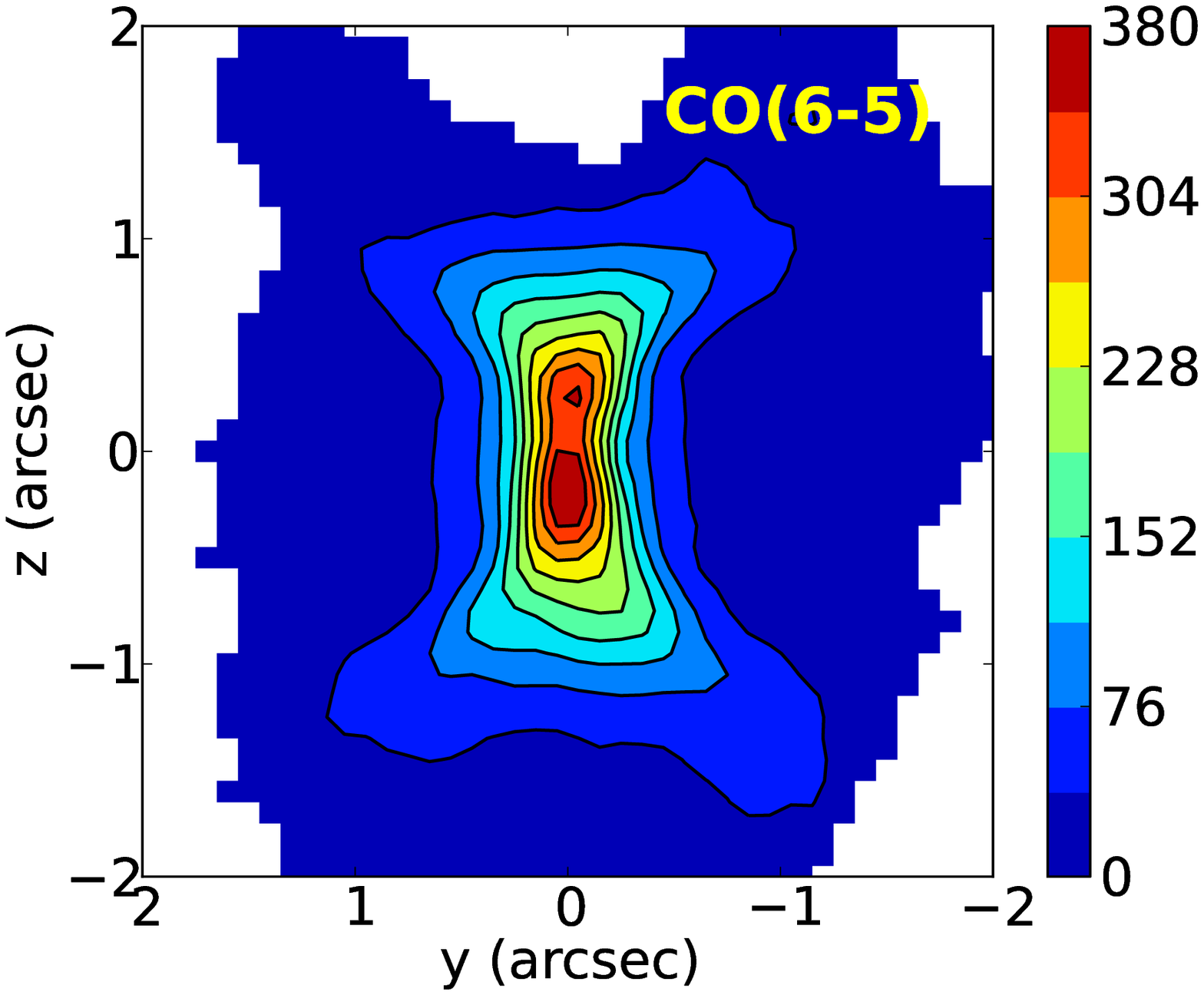} 
\end{tabular}
\caption{From left to right: 1) HST wide field planetary camera 2 image of the Red Rectangle from Cohen et al. (2004). 2) Keck telescope near-infrared speckle image from Tuthill et al. (2002). North is up and East is left. 3) and 4) CO images (4$''$$\times$4$''$) rotated by 13$^\circ$ clockwise integrated over Doppler velocities from $-$7.2 kms$^{-1}$ to 7.2 kms$^{-1}$ (present work).}\label{Fig1}
\end{center}
\end{figure*}

The Red Rectangle was first described in detail by Cohen et al. (1975) and since then has been the target of many observations and the object of numerous modelling exercises. It owes its name to its appearance in the visible, as observed recently by Cohen et al. (2004) using the Hubble Space Telescope and shown in Figure \ref{Fig1}. It is commonly accepted that this appearance is the result of a biconical structure with axis perpendicular to the line of sight. The star in the centre, HD 44179, is known to be a binary made of a post AGB star and a secondary star accreting the wind of the former in a disk perpendicular to the star axis and therefore to the sky plane. The binary is an unresolved spectroscopic binary, the main parameters of which are well known. The secondary star is usually considered to be a low mass main sequence star but some authors argue in favour of a white dwarf (see for example Men'shchikov et al. 2002). The general idea is that a fast jet normal to the accretion disk has dug a conical cavity in the slow wind of the post AGB star. The appearance at visible wavelengths is complicated by the fact that the light observed is the light emitted by the star and diffused on the walls of the conical cavities, direct light being prevented to reach the Earth by the presence of a dense dust torus around the star. Several authors have proposed models along such lines. Two recent examples from which earlier references can easily be traced are Koning et al. (2011) and Thomas (2012).
An open question is the nature of the bipolar outflow carving the conical cavities in the slowly expanding wind of the post-AGB star. While it is commonly accepted that it has a high velocity, in excess of \mbox{158 kms$^{-1}$} according to Koning et al. (2011), its opening angle might be broad, typically at the scale of the conical cavity, or narrow but precessing around the star axis. A recent polarization measurement (Martinez Gonzalez et al. 2014) pleads in favour of the second hypothesis.

Until recently, radio and millimetre wave observations of the Red Rectangle had insufficient resolution to map the morphology of molecular gas emission. The only exception was a Plateau de Bure IRAM observation of CO(2-1) and CO(1-0) emission, with a resolution of $\sim$1$''$, sufficient to reveal the presence of a disk in rotation perpendicular to the star axis (Bujarrabal et al. 2005). Using the experience gained by observing other proto-planetary nebulae, the detailed study of the profiles of other rotational lines confirmed the presence of such a disk (Bujarrabal et al. 2013, 2013a). Recently, ALMA observations in \mbox{CO(6-5)} and \mbox{CO(3-2)} with an order of magnitude better resolution (Bujarrabal et al. 2013b) have been made available. Their analysis is the object of the present work.

\begin{figure*}[ht]
\begin{center}
\includegraphics[scale=0.24,trim=0.2cm 0.2cm 0.2cm 0.2cm,clip]{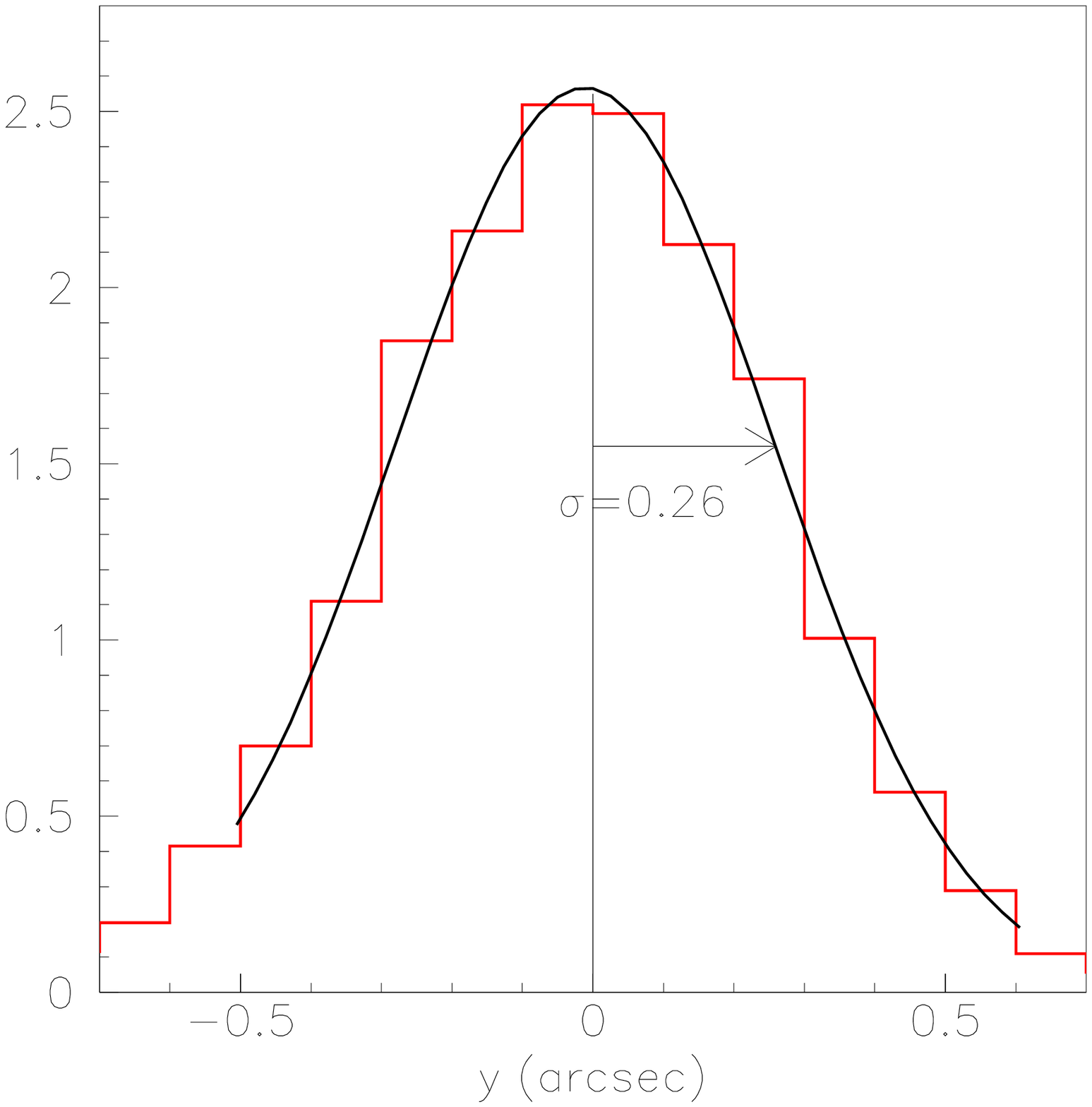}
\includegraphics[scale=0.24,trim=0.2cm 0.2cm 0.2cm 0.2cm,clip]{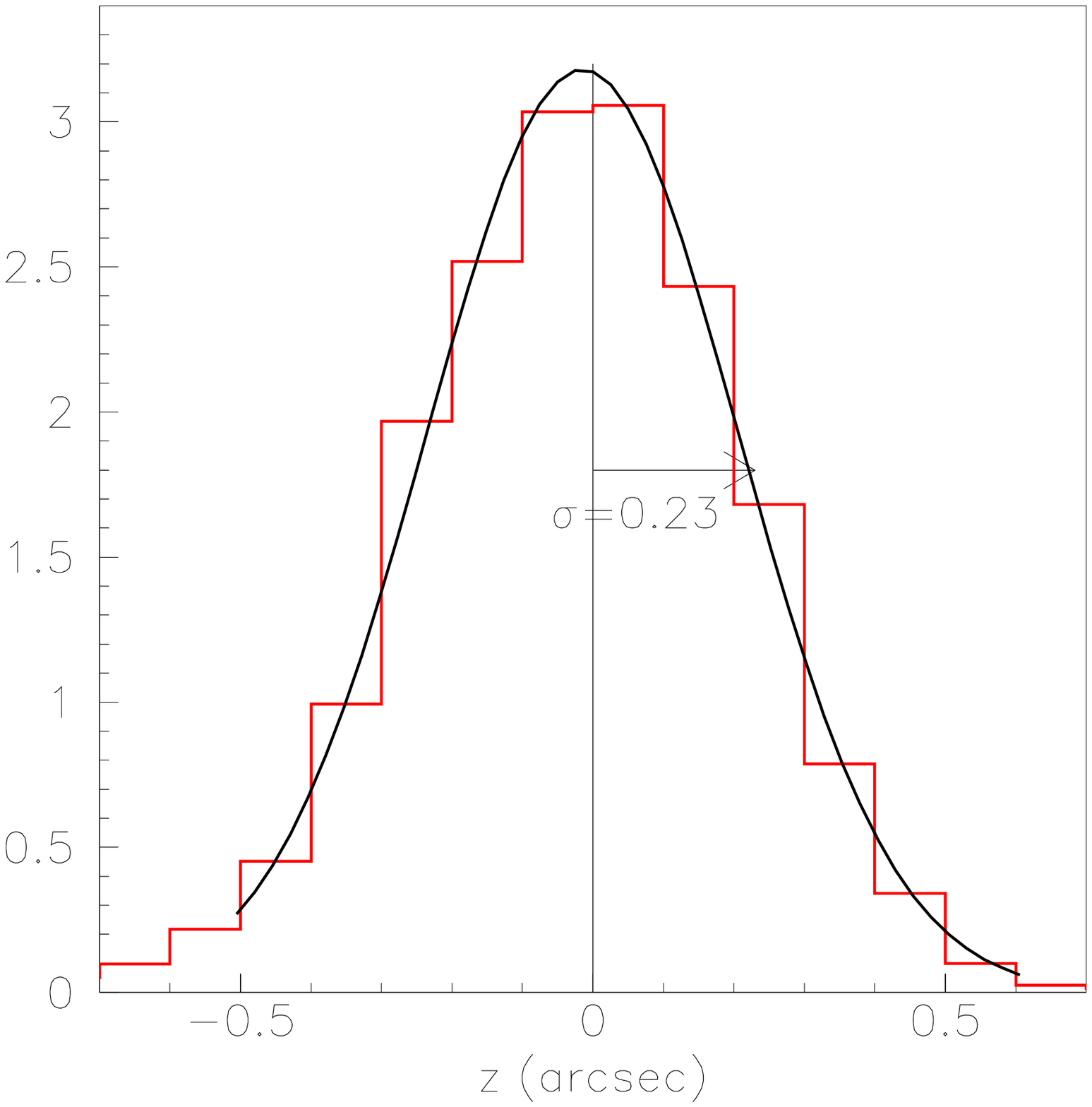}
\includegraphics[scale=0.24,trim=0.2cm 0.2cm 0.2cm 0.2cm,clip]{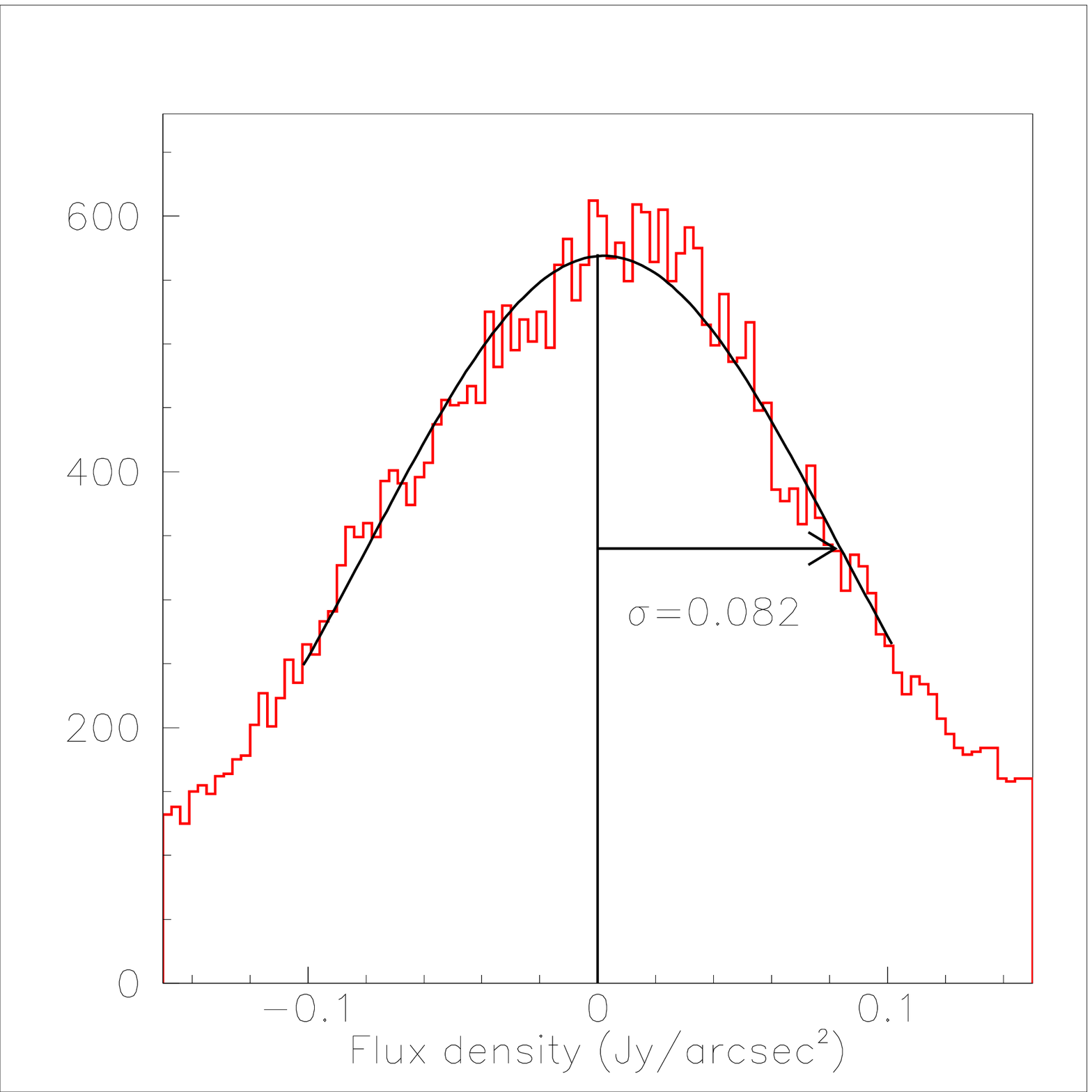}
\includegraphics[scale=0.24,trim=0.2cm 0.2cm 0.2cm 0.2cm,clip]{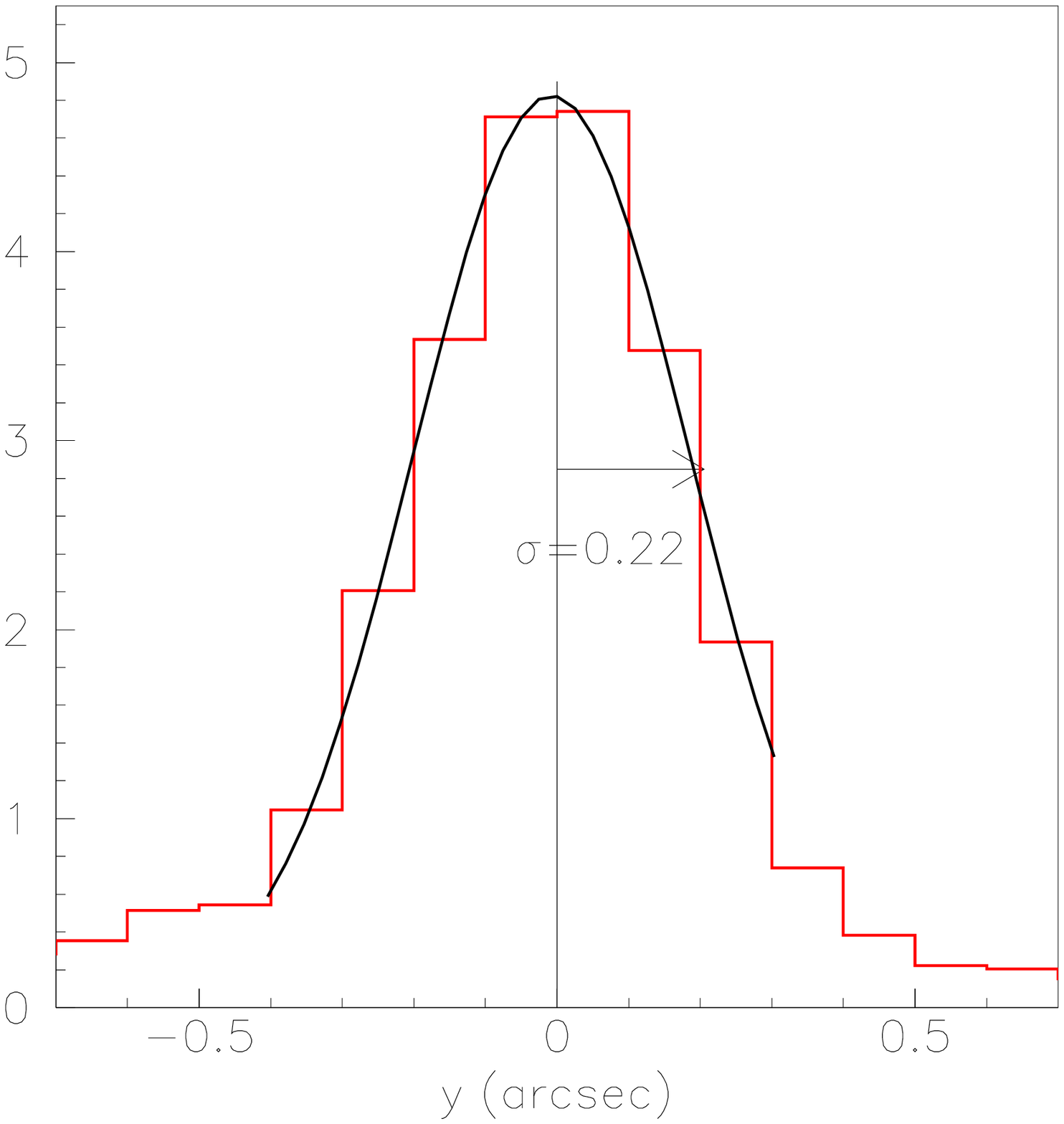}
\includegraphics[scale=0.24,trim=0.2cm 0.2cm 0.2cm 0.2cm,clip]{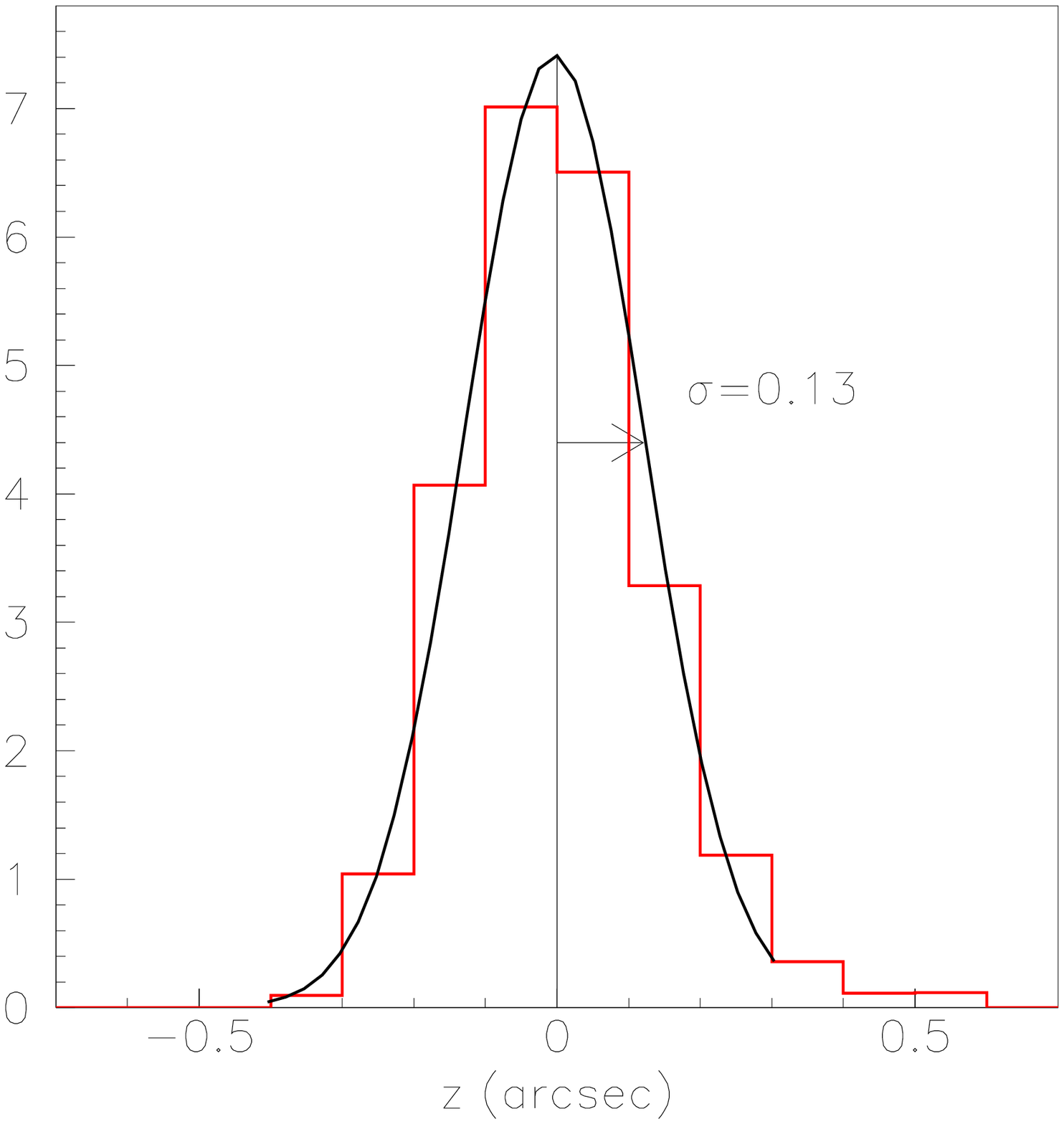}
\includegraphics[scale=0.24,trim=0.2cm 0.2cm 0.2cm 0.2cm,clip]{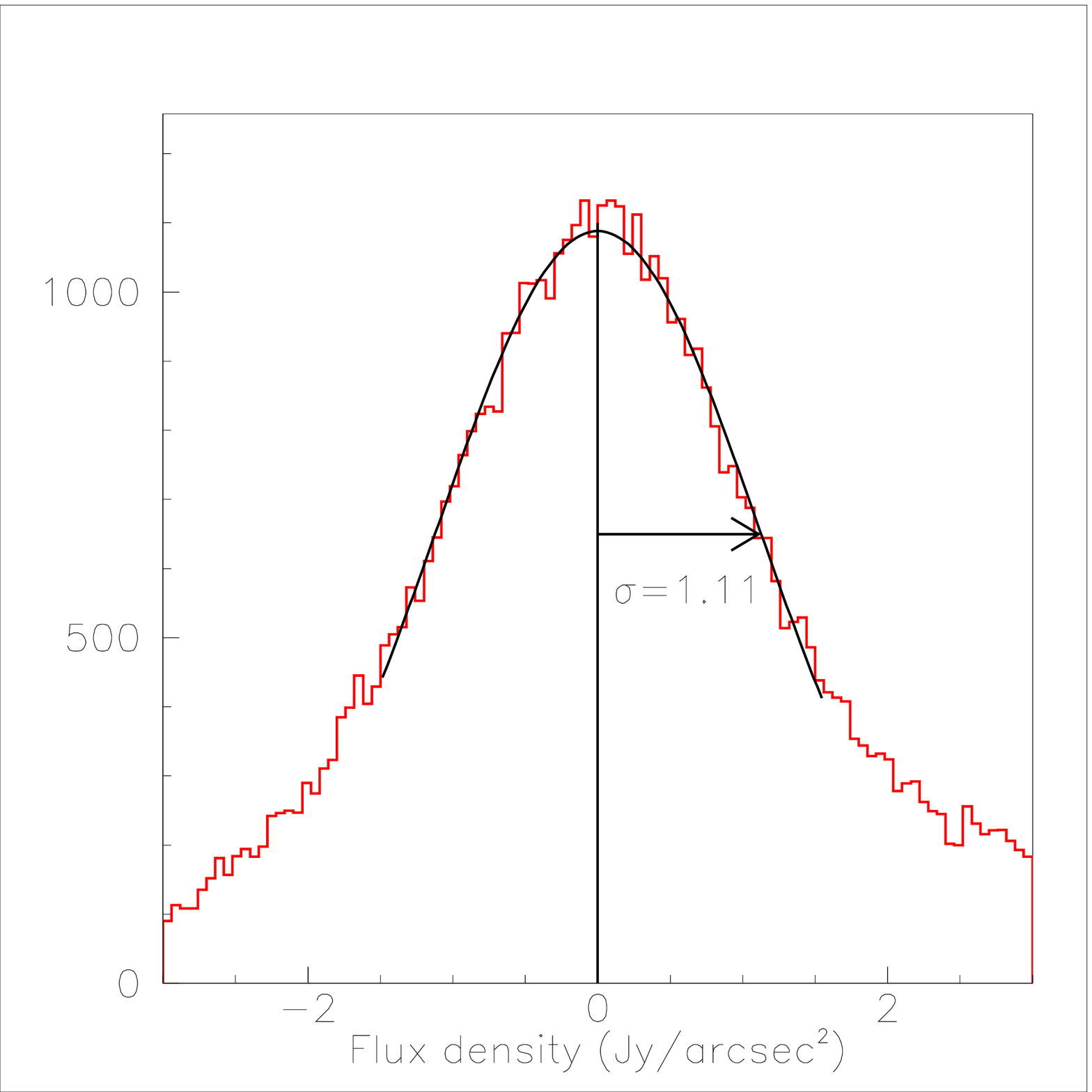}
\caption{Projections on $y$ (left) and $z$ (middle) of the continuum emission (Jy/beam). Right:  Lower ends of the line flux density distributions (Jy$\times$arcsec$^{-2}$). The upper panels are for \mbox{CO(3-2)} and the lower panels for \mbox{CO(6-5)}. Gaussian fits are shown on the peaks.}\label{Fig2}
\end{center}
\end{figure*}
\section{DATA}
\label{sec:data}
\mbox{CO(3-2)} and \mbox{CO(6-5)} ALMA data of the Red Rectangle (RR) have been recently released for public access (Project number 2011.0.00223.S). \mbox{CO(3-2)} data are available in the form of data-cubes of 360$\times$360 pixels and 70 frequency bins, \mbox{CO(6-5)} data of 432$\times$432 pixels and 43 frequency bins. For convenience, we have rearranged both \mbox{CO(3-2)} and \mbox{CO(6-5)} data in a common array of 50$\times$50 pixels centred on the continuum emission of the star and covering a solid angle of 5$''$$\times$5$''$,with Doppler velocity spectra covering from $-$7.2 kms$^{-1}$ to 7.2 kms$^{-1}$ in 36 bins of 0.40 kms$^{-1}$ each. Moreover, we have rotated the new array by 13$^\circ$ counter clockwise in order to have it aligned with the star axis. The value of 13$^\circ$ used for the position angle of the star axis is consistent with values found in the literature (between 10$^\circ$ and 15$^\circ$) and with a preliminary analysis which we made of the data, giving as result 13$^\circ$$\pm$2$^\circ$. Both the sky maps and the velocity spectra associated with the new arrays are centred on the star, the former using the continuum data with a precision of $\sim$0.02$''$ and the latter such that the mean Doppler velocity evaluated over pixels distant by less than 2.5$''$ from the star cancels, with a precision of $\sim$0.1 kms$^{-1}$. Details of data collection are given in Appendix A of Bujarrabal et al. (2013b) and do not need to be repeated here, where we use the data in the form provided by the ALMA staff, which are of a quality sufficient for the purpose of the present study. The synthetic beam sizes are 0.50$''$$\times$0.49$''$ and 0.27$''$$\times$0.24$''$ for \mbox{CO(3-2)} and \mbox{CO(6-5)} respectively.

\begin{figure*}[ht]
\begin{center}
\includegraphics[scale=0.23,trim=2.0cm 0. -0.5cm 0.,clip]{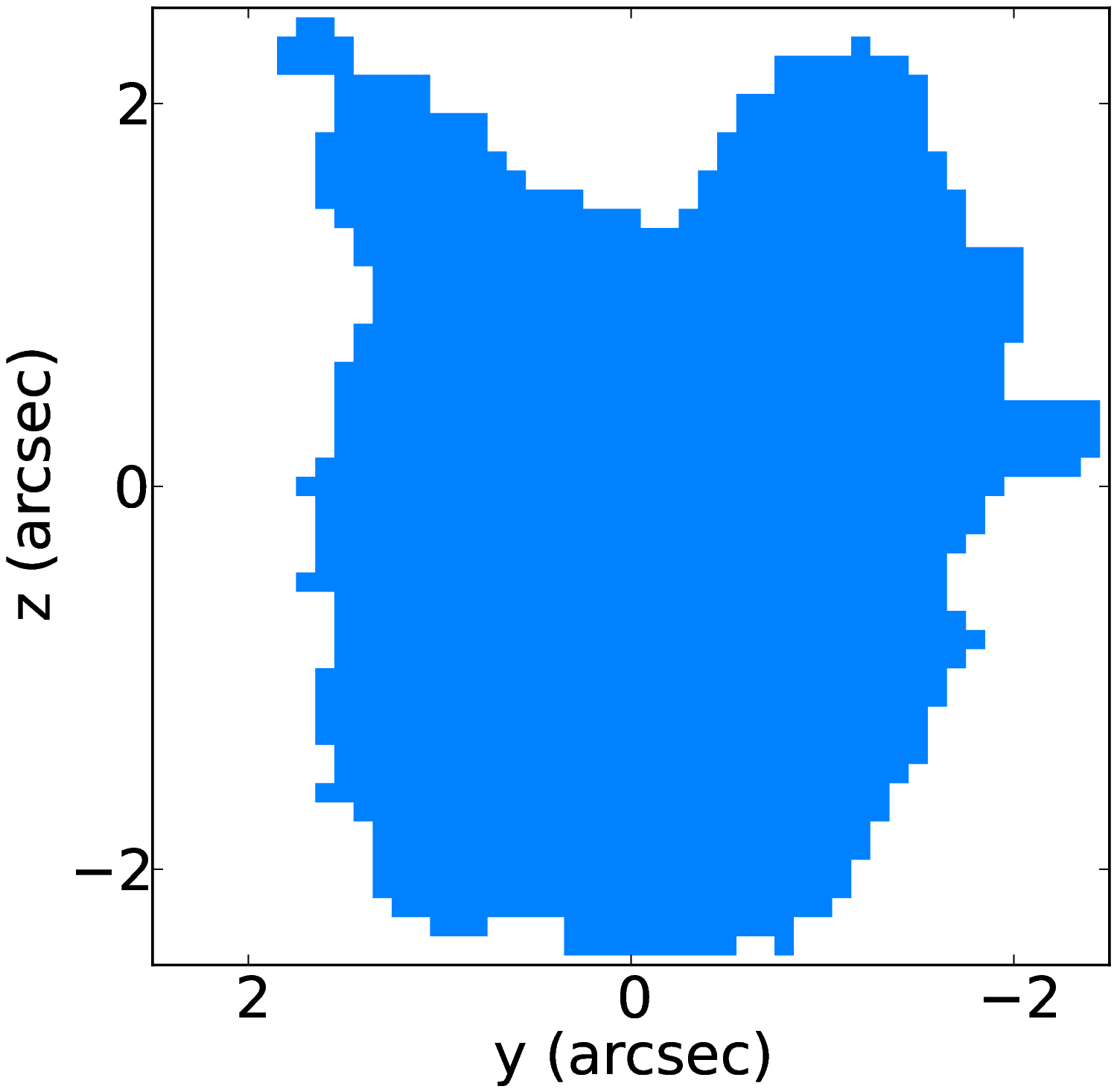}
\includegraphics[scale=0.23,trim=.5cm 0. 0.5cm 0.,clip]{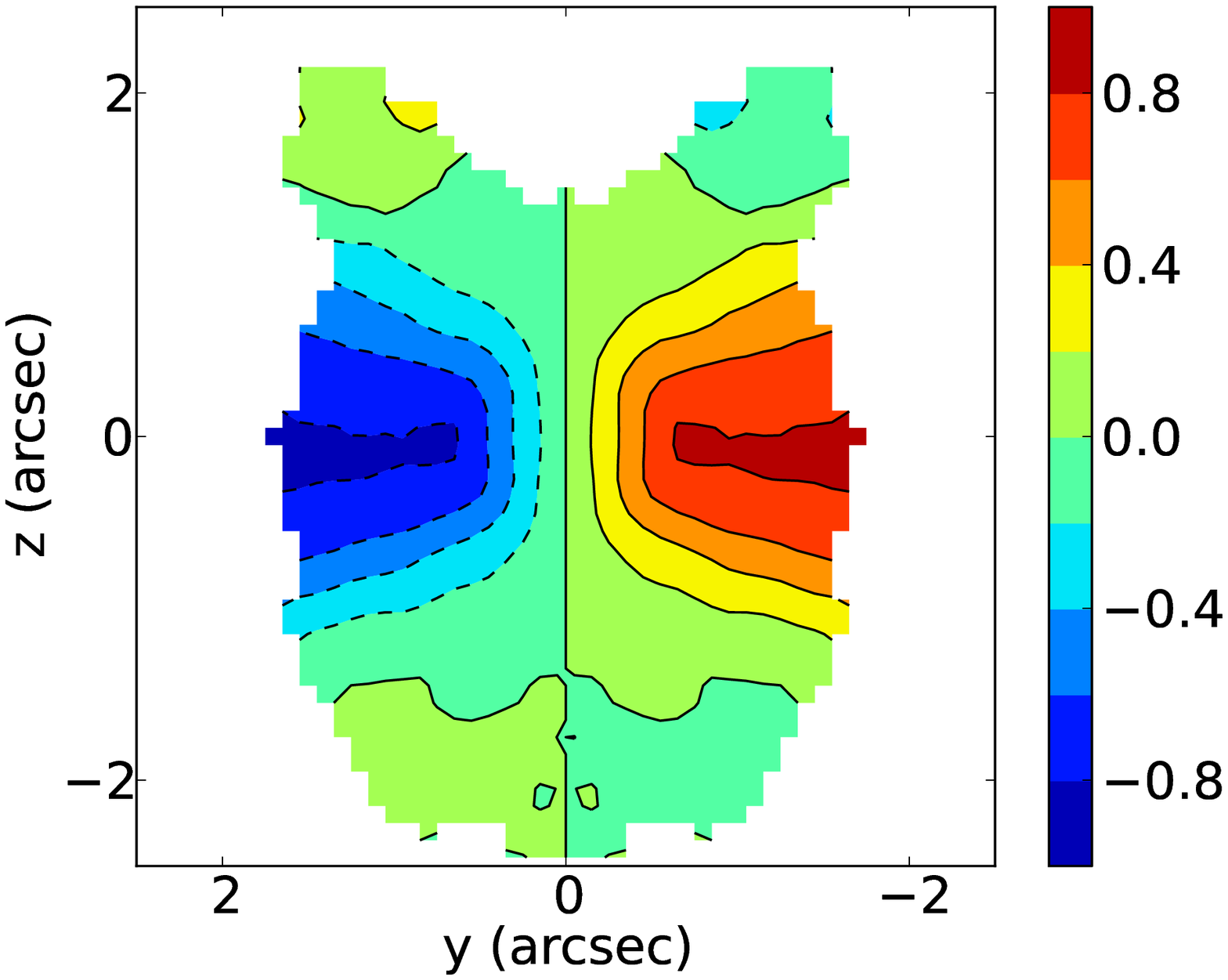}
\includegraphics[scale=0.23,trim=.5cm 0. 0.5cm 0.,clip]{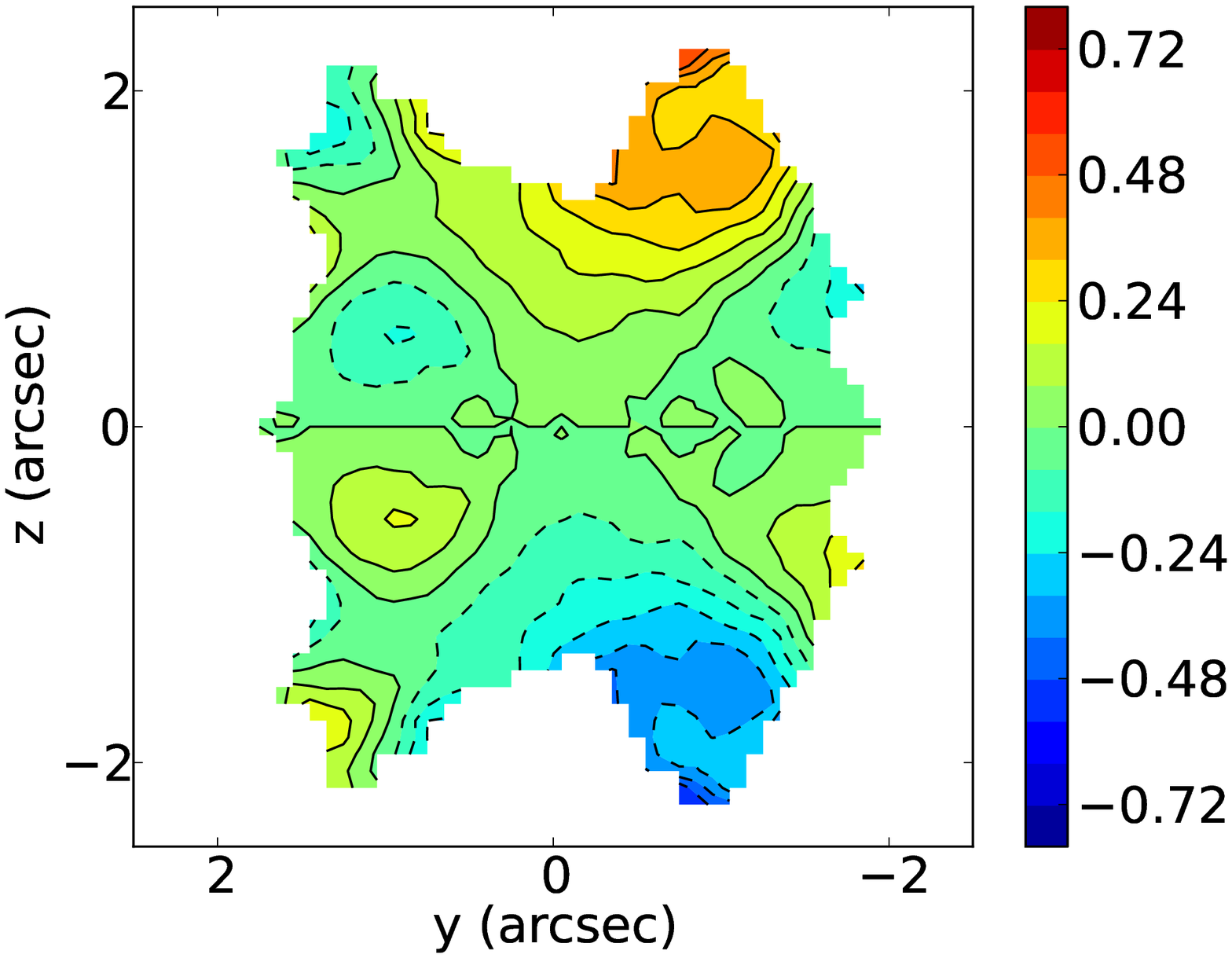}
\includegraphics[scale=0.23,trim=.0cm 0. 0.5cm 0.,clip]{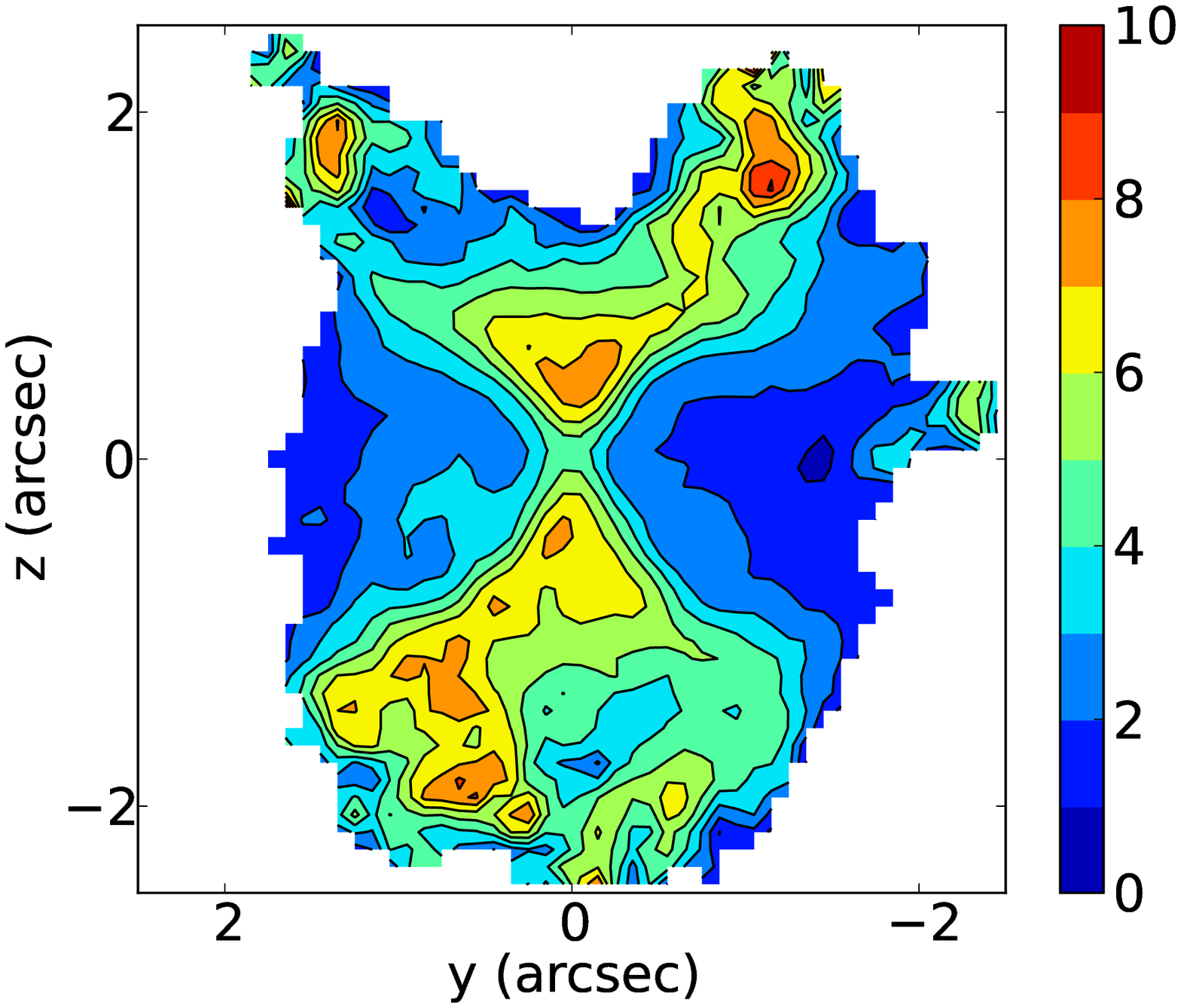}
\includegraphics[scale=0.23,trim=.5cm 0. 0.5cm 0.,clip]{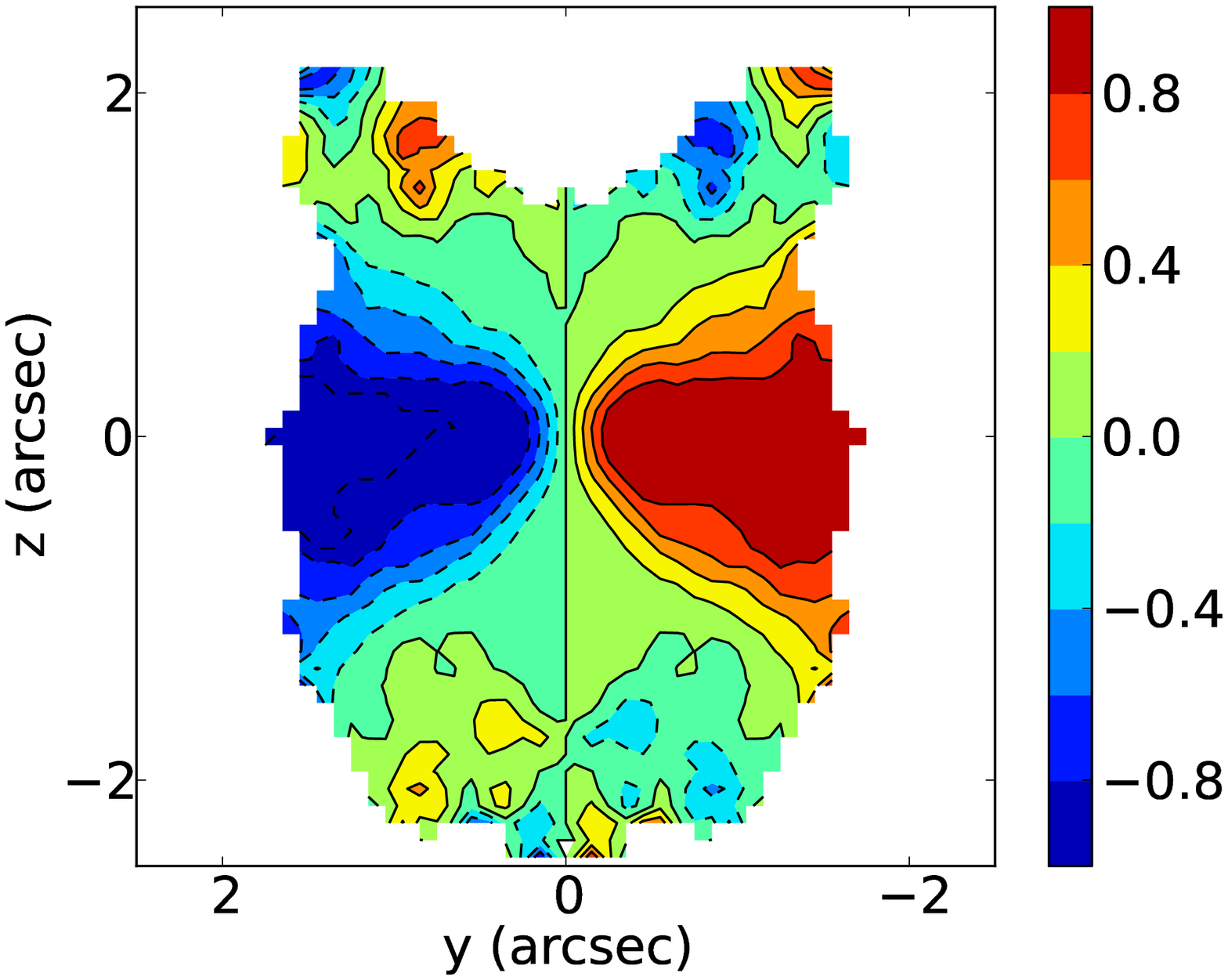}
\includegraphics[scale=0.23,trim=.5cm 0. 0.5cm 0.,clip]{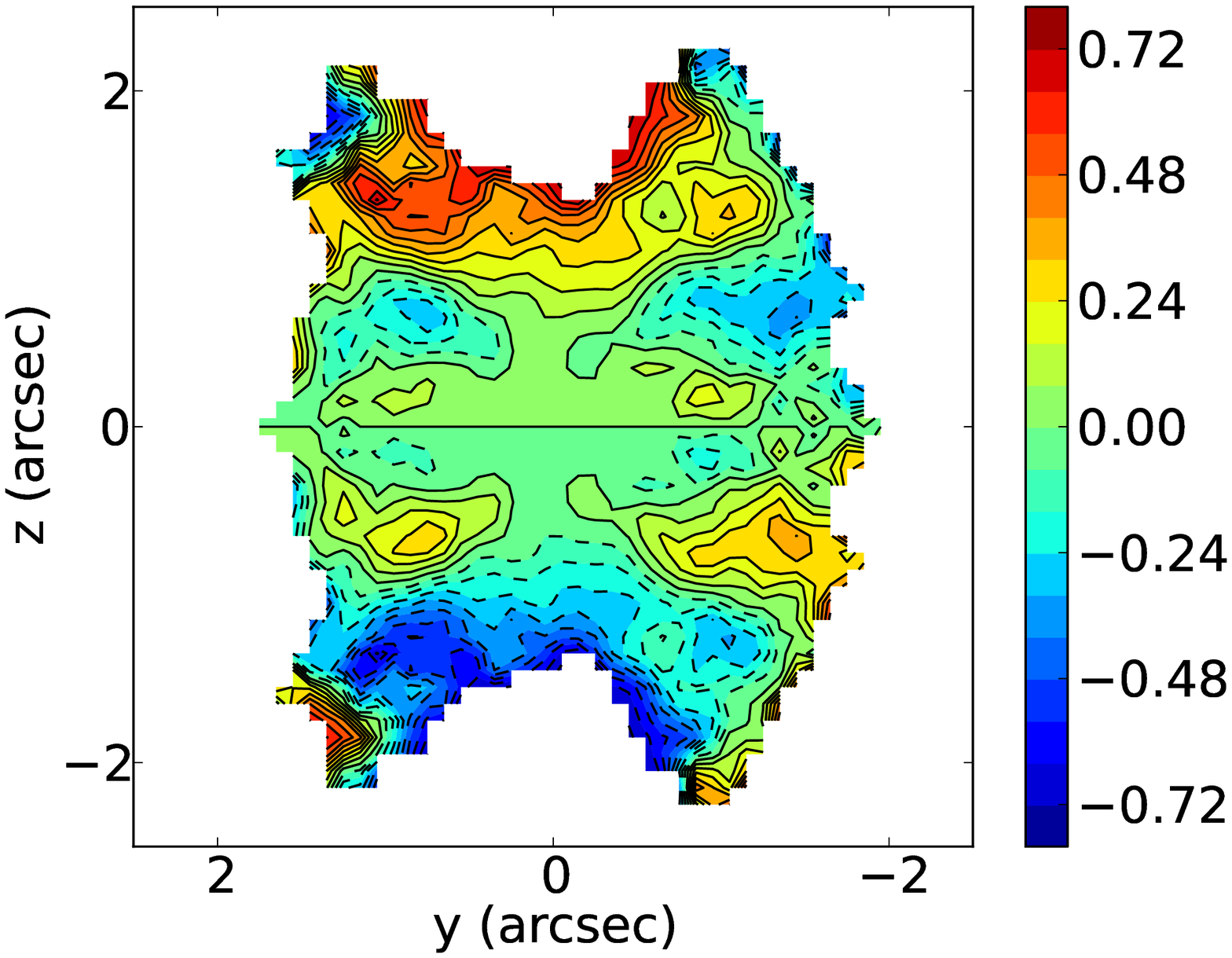}
\caption{Left: the upper panel displays (in blue) the map of retained pixels and the lower panel that of $A_T$, the \mbox{CO(6-5)} to \mbox{CO(3-2)} flux ratio. Sky maps of $A_\eta$ (middle) and $A_z$ (right) are shown for \mbox{CO(3-2)} and \mbox{CO(6-5)} in the upper and lower panels respectively.}\label{Fig3}
\end{center}
\end{figure*}

Figure \ref{Fig2} (left and middle) shows the projections of the continuum emission on the axes of the rotated array, used to centre the sky map; continuum emission is discussed in Section 8. Figure \ref{Fig2} (right) displays the lower ends of the line flux density distributions; after small adjustment of the empty sky baselines, Gaussian fits give rms values of 20 mJy/beam for \mbox{CO(3-2)} and 72 mJy/beam for \mbox{CO(6-5)}, corresponding respectively to 0.082 and 1.11 Jy$\times$arcsec$^{-2}$. The image reprocessing made by Bujarrabal et al. (2013b) yields significantly lower values (respectively 7 and 30 mJy/beam) with the implication that most of the flux fluctuations observed here arise from imperfect deconvolution and/or calibration. As a consequence, the uncertainty attached to the sum of $N$ data cubes is far from being equal to the above values multiplied by $\sqrt{N}$ and must be evaluated for each specific case separately. In particular, when integrating pixel flux densities over velocity spectra we find that reliable results are obtained by retaining only pixels containing more than 0.75 Jy$\times$kms$^{-1}$$\times$arcsec$^{-2}$ in the \mbox{CO(3-2)} data sample and more than 3.0 Jy$\times$kms$^{-1}$$\times$arcsec$^{-2}$ in the \mbox{CO(6-5)} data sample. The map covered by the retained pixels is illustrated in the upper left panel of Figure \ref{Fig3} and contains 1526 pixels (covering 15.3 arcsec$^{2}$). The restriction to such a region of the sky is conservative but sufficient for the purpose of the present study. However, improved deconvolution and calibration, such as made by Bujarrabal et al. (2013b), would allow for exploring finer details, in particular at large distance from the star.

$^{13}$CO data are also available from the same ALMA observations but have not been considered in the present work.

\section{MAIN FEATURES}
\label{sec:features}
 We present below general information that can be obtained from the data without having recourse to a model. We pay particular attention to assessing the symmetries of the structure and to comparing \mbox{CO(6-5)} and \mbox{CO(3-2)} emissions (an indicator of the temperature when absorption can be neglected). Figure \ref{Fig1} shows the sky maps of \mbox{CO(3-2)} and \mbox{CO(6-5)} emissions integrated over the lines. They display the same symmetries as the visible and infrared maps.

We use coordinates $x$, $y$ and $z$, respectively pointing away from Earth (along the line of sight), 13$^\circ$ south of East and 13$^\circ$ east of North (along the star axis), all measured as angular distances in arcseconds (the star is usually considered as being 710 pc away from Earth, see Men'shchikov et al. 2002 ). We define a space radius $r=\sqrt{x^2+y^2+z^2}$, a sky radius $R=\sqrt{y^2+z^2}$ and an equatorial radius $\xi=\sqrt{x^2+y^2}$. To a good approximation, the inclination angle of the star axis with respect to the sky plane, $\theta$, is known to cancel. Under such an approximation, symmetry properties of the gas morphology and kinematics take remarkably simple forms in terms of the measured flux densities, $F(y,z,V_x)$, $V_x$ being the Doppler velocity.
\begin{figure*}[ht]
\begin{center}
\includegraphics[trim = 0cm .8cm 0cm 0cm, clip, scale=0.72]{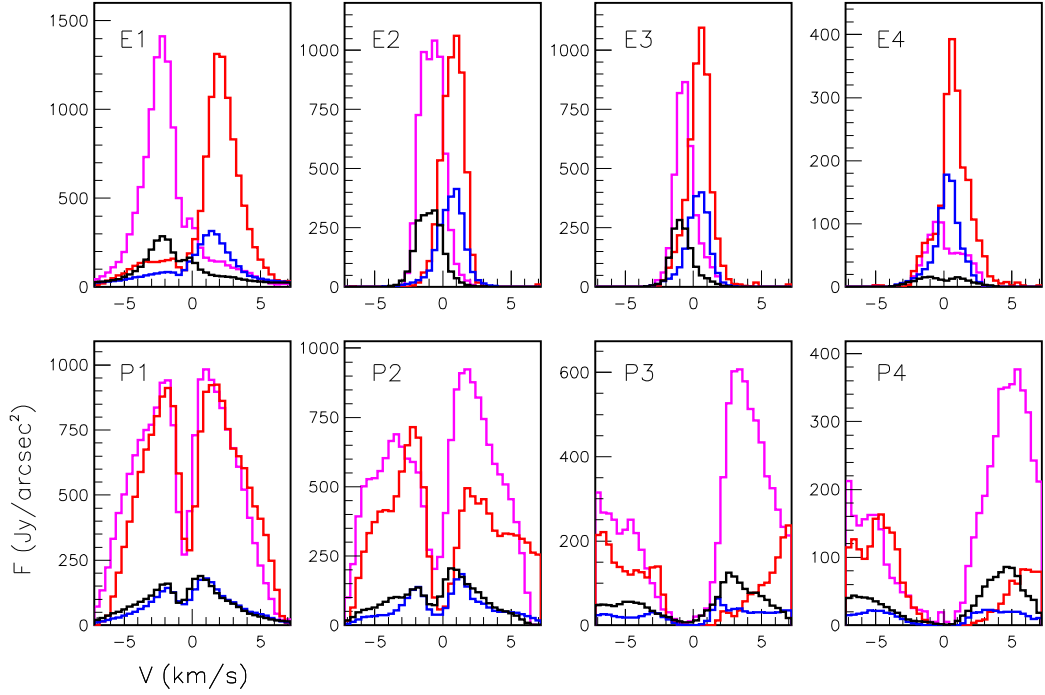}
\caption{Velocity spectra integrated over sky regions defined in the text. Red and magenta is for \mbox{CO(6-5)}, blue and black for \mbox{CO(3-2)}. P and E stand for polar and equatorial sectors respectively, followed by a digit labelling the radial rings from centre outward. In the upper panels, magenta and black are for eastern regions, blue and red for western regions. In the lower panels, red and blue are for northern regions and black and magenta for southern regions.}\label{Fig4}
\end{center}
\end{figure*}
We define star cylindrical coordinates, $(\xi,\omega,z)$ with $\xi=x/sin\omega$ and $tan\omega=-x/y$. Introducing a coordinate $\eta=xcos\omega+ysin\omega$, normal to the meridian plane, the components $V_{\xi}$, $V_{\eta}$ and $V_z$ of the space velocity contribute respectively $V_{\xi}sin\omega$, $V_{\eta}cos\omega$ and 0 to the Doppler velocity. Moreover, relaxing the condition $\theta$=0 simply introduces, to first order in $\theta$, a contribution $\theta V_z$. The Doppler velocity reads then:
\begin{equation}{\label{eq1}} 
V_x=(x/\xi)V_{\xi}-(y/\xi)V_{\eta}+\theta V_z
\end{equation}
 It is convenient to introduce an effective density $\rho(x,y,z)$ such that the flux density integrated over Doppler velocities observed at a point ($y,z$) in the sky plane reads
\begin{equation}{\label{eq2}} 
F(y,z)=\int{F(y,z,V_x)dV_x}=\int{\rho(x,y,z)dx}
\end{equation}
the second integral being taken along the line of sight. When discussing symmetries, it is also convenient to introduce blue shifted and red shifted components, $B$ and $R$ \mbox{($F=B+R$)}, with the the first integral running over \mbox{$V_x<0$} and \mbox{$V_x>0$} respectively. Under the hypothesis that the properties of the star are invariant by rotation about the star axis (namely independent from $\omega$), at a given $z$, $V_{\xi}$ contributes \mbox{$R(y)=B(y)=R(-y)=B(-y)$} and $V_{\eta}$ contributes \mbox{$R(y)=B(-y), B(y)=R(-y)$}. A good indicator of rotation about the star axis is therefore the quantity $A_{\eta}=$\mbox{$[R(y)+B(-y)-R(-y)-B(y)]/[F(y)+F(-y)]$}. Figure \ref{Fig3} displays maps on the sky plane of \mbox{$A_T=F(CO[6-5])/F(CO[3-2])$}, $A_{\eta}$ and the North-South asymmetry, $A_z=[F(y,z)-F(y,-z)]/[F(y,z)+F(y,-z)]$.

 The first of these quantities is sensitive to temperature. Indeed, in addition to the actual CO density, namely gas density multiplied by CO abundance, the effective density introduced in Relation 2 accounts for the population of the excited molecular level and the emission probability. Under the hypothesis of local thermal equilibrium, it depends therefore essentially on temperature. The assumption of rotational invariance about the star axis implies in addition that absorption is negligible. Integrating the flux over the pixels that have been retained, we obtain respectively 142 Jy$\times$kms$^{-1}$ and 615 Jy$\times$kms$^{-1}$ for \mbox{CO(3-2)} and \mbox{CO(6-5)}. Under the hypothesis of thermal equilibrium and neglecting absorption, their ratio, 4.3, corresponds to an average temperature of 63 K. A detailed discussion of the distribution of the gas temperature is given in Section 5. The $A_T$ map reveals the biconical structure in a very clear way, providing evidence for a temperature distribution dominated by the morphology of the outflow down to small distances to the star. It also reveals inhomogeneities of the biconical outflow that are discussed in Section 7.
The second quantity, $A_{\eta}$, reaches very high values and provides spectacular evidence for rotation over a broad angular range about the equator, the eastern part being blue-shifted and the western part red-shifted. There is no sign of a thin equatorial disk.

The third quantity, $A_z$, displays a significant north-south asymmetry of the bipolar outflow at large distances from the star, however much smaller than in the case of $A_{\eta}$.

 A summary of the main features of the observed morphology and kinematics is presented in Figure \ref{Fig4} where each of the four quadrants of the sky plane, North-East, North-West, South-East and South-West, are folded together and segmented in four radial and two angular intervals. The radial intervals (sky radius $R$) are 0.6$''$ wide starting at 0.1$''$. The angular regions are a polar sector for position angles inferior to 32.5$^\circ$ with respect to the star axis ($z$) and an equatorial sector for position angles inferior to 47.5$^\circ$ with respect to the equator ($y$). Figure \ref{Fig4} displays velocity spectra integrated over the eight regions defined by this ring and sector geometry. The equatorial region is dominated by rotation, with velocities decreasing with $R$ from $\sim$2 kms$^{-1}$ at $R$$\sim$0.5$''$ to $\sim$0.7 kms$^{-1}$ at $R$$\sim$1.5$''$. The polar regions are instead dominated by an outflow, with Doppler velocities extending to, and even beyond, the limits of the spectra. The outflow velocity is seen to increase significantly with $R$. It may be due to the presence of a velocity gradient, the gas being accelerated continuously over the whole range of $R$ explored here, or to an opening of the biconical cavity when $R$ increases. As was already remarked in Figure \ref{Fig3}, strong deviations from symmetry are present at large values of $R$, in particular in the \mbox{CO(6-5)} data.

\section{GAS EFFECTIVE DENSITY}
\label{sec:dens}
 The complexity of the RR morphology is an invitation to proceed by steps in the evaluation of its geometrical, kinematical and physical properties.

 In a first step we evaluate the effective densities from the values taken by the measured flux densities integrated over the velocity spectra for \mbox{CO(3-2)} and \mbox{CO(6-5)} separately. We assume that the effective densities obey rotational symmetry about the star axis, namely that they are functions of $\xi$ and $z$ exclusively. The study of the effect of such an assumption is kept for the last step. The effective densities are obtained by simply solving the integral equation $F(y,z)=\int\rho(\xi,z)dx$ (Relation 2).
 In a second step, retaining the assumption of rotational symmetry, we evaluate the field of gas velocities using the values of the effective densities obtained in the first step and fit the \mbox{CO(3-2)} and \mbox{CO(6-5)} data together with a same velocity distribution. Here, the $\xi$ (expansion) and $\eta$ (rotation) components of the velocity vectors need to be evaluated as functions of $\xi$ and $z$, implying the adoption of a model, which we choose as simple as possible.

 The last step studies the asymmetries that have been neglected in the first two steps and the implications of the assumptions that have been made.
\begin{figure*}[htb]
\begin{center}
\includegraphics[scale=0.35]{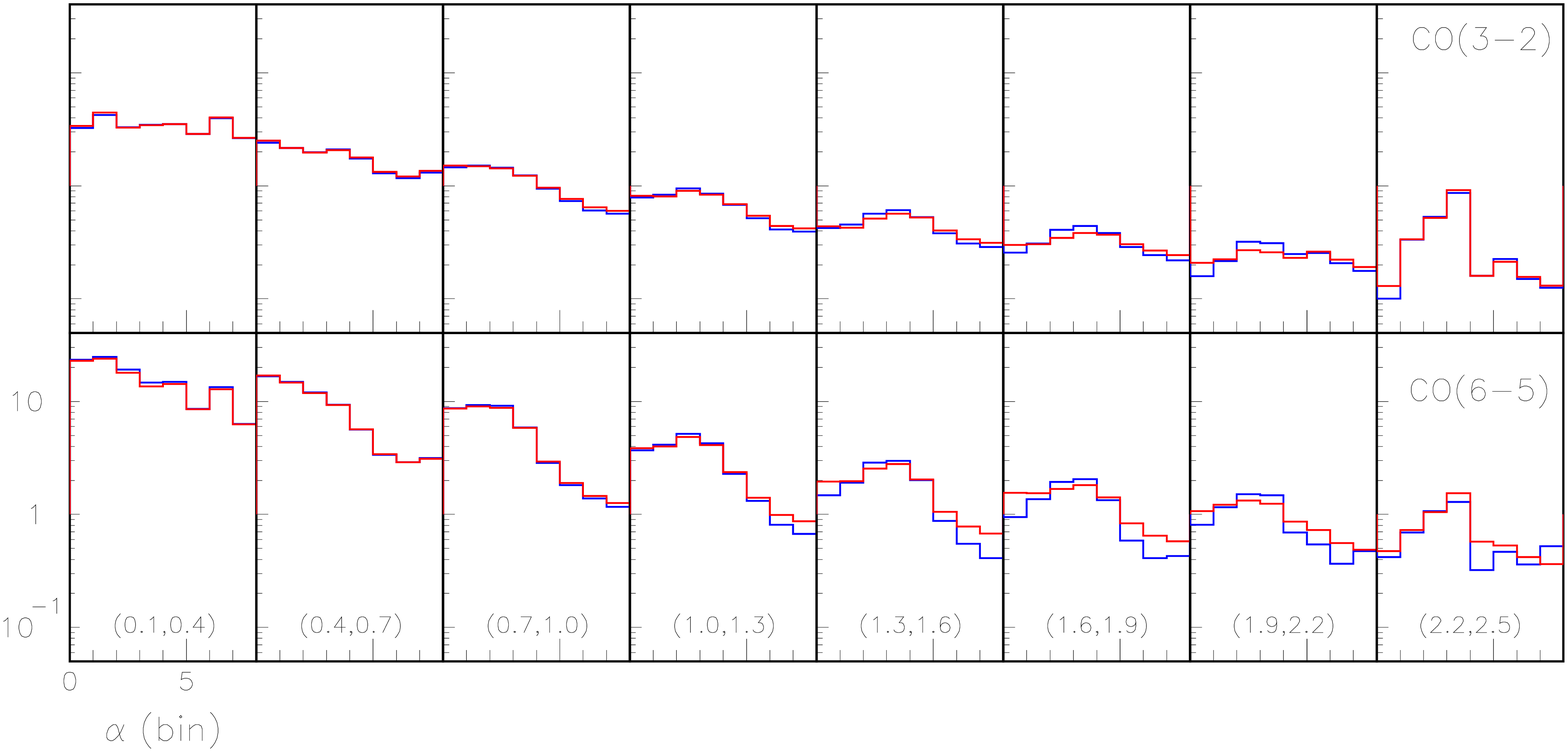}
\includegraphics[scale=0.35]{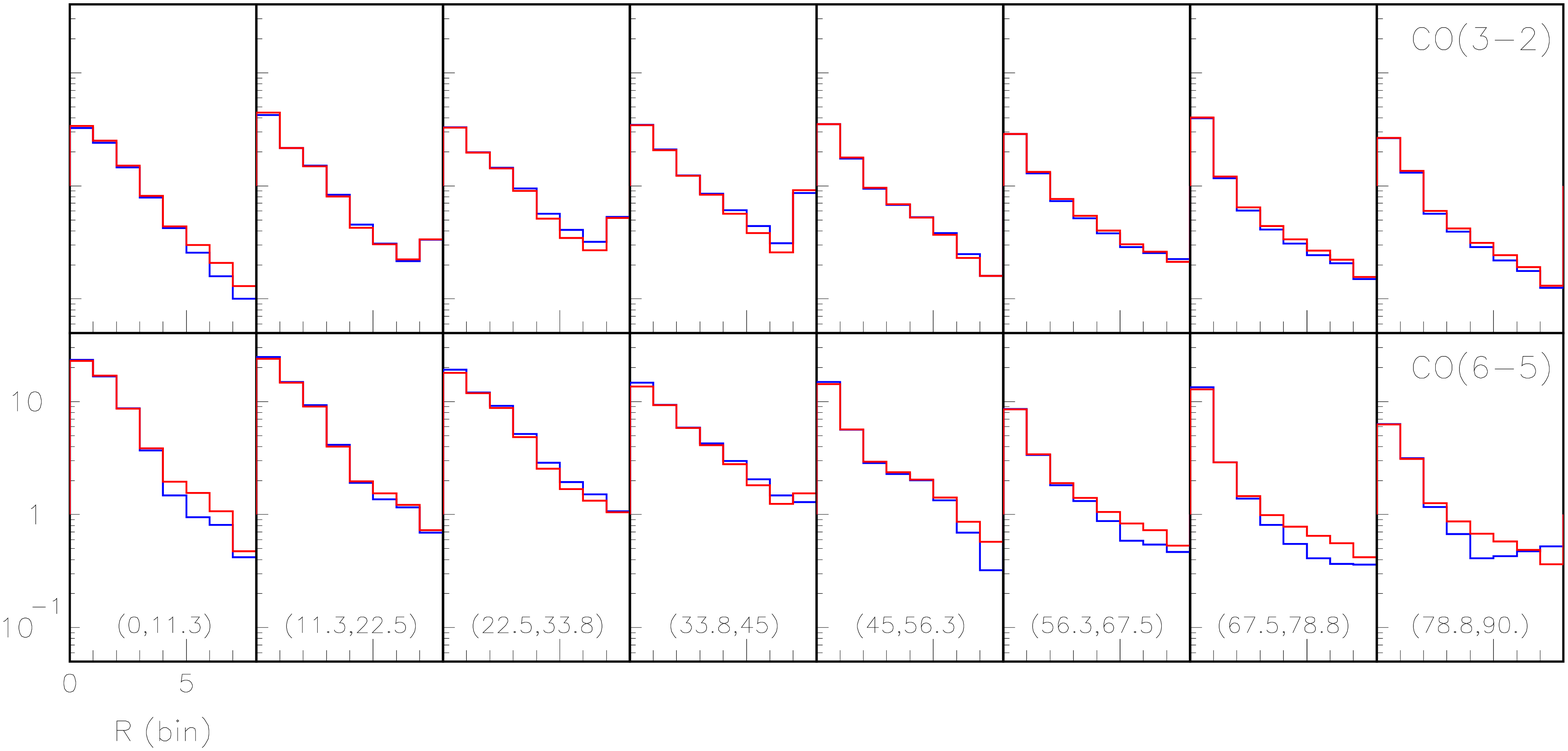}
\caption{Measured flux densities (blue) averaged over ($R$,$\alpha$) intervals of sizes (0.3$''$,11.25$^\circ$) are compared with the result (red) of integrating over the line of sight the effective densities obtained by solving the integral equation. The upper panels display $\alpha$ distributions in eight successive $R$ intervals, the lower panels display $R$ distributions in eight successive $\alpha$ intervals. In each case, the upper row is for \mbox{CO(3-2)} and the lower row for \mbox{CO(6-5)}. Panels are labelled with the corresponding interval, in arcseconds for $R$ and degrees for $\alpha$.}\label{Fig5}
\end{center}
\end{figure*}
\begin{figure*}[htb]
\begin{center}
\includegraphics[trim= 1.5cm 1.cm .5cm 1.cm, clip, scale=0.34]{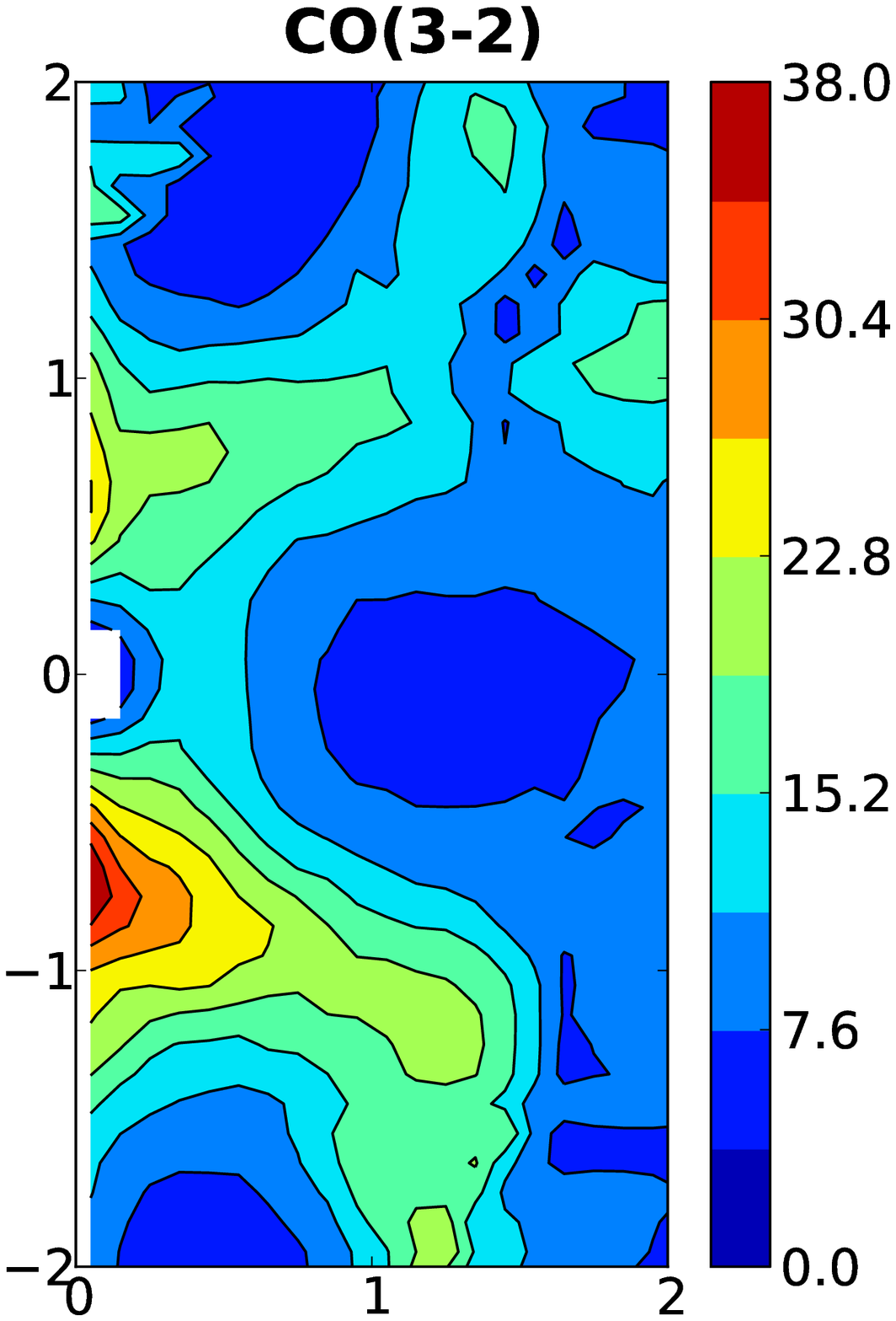} 
\includegraphics[trim= 1.5cm 1.cm .5cm 1.cm, clip, scale=0.34]{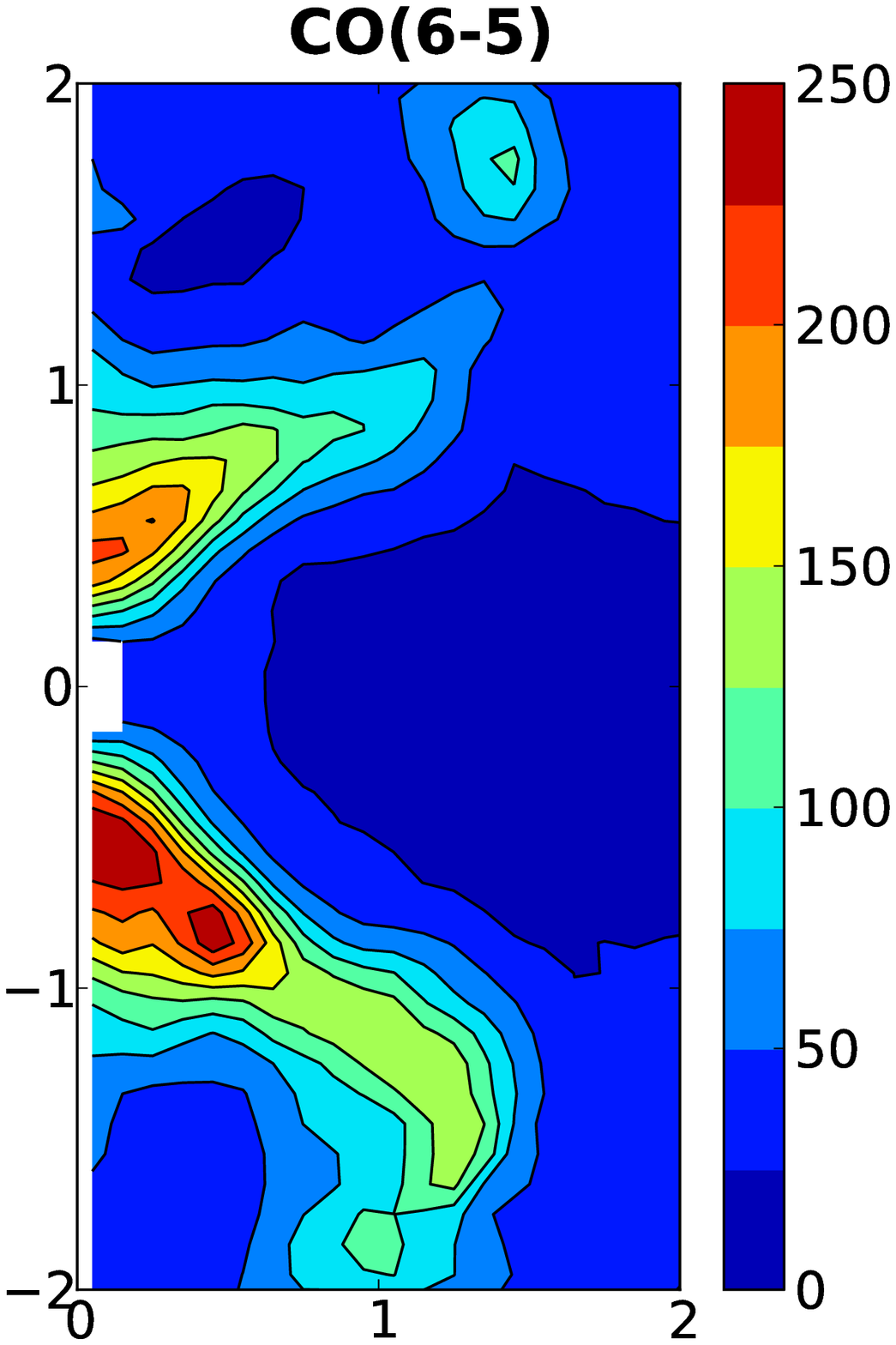} 
\includegraphics[trim= 1.5cm 1.cm .5cm 1.cm, clip, scale=0.34]{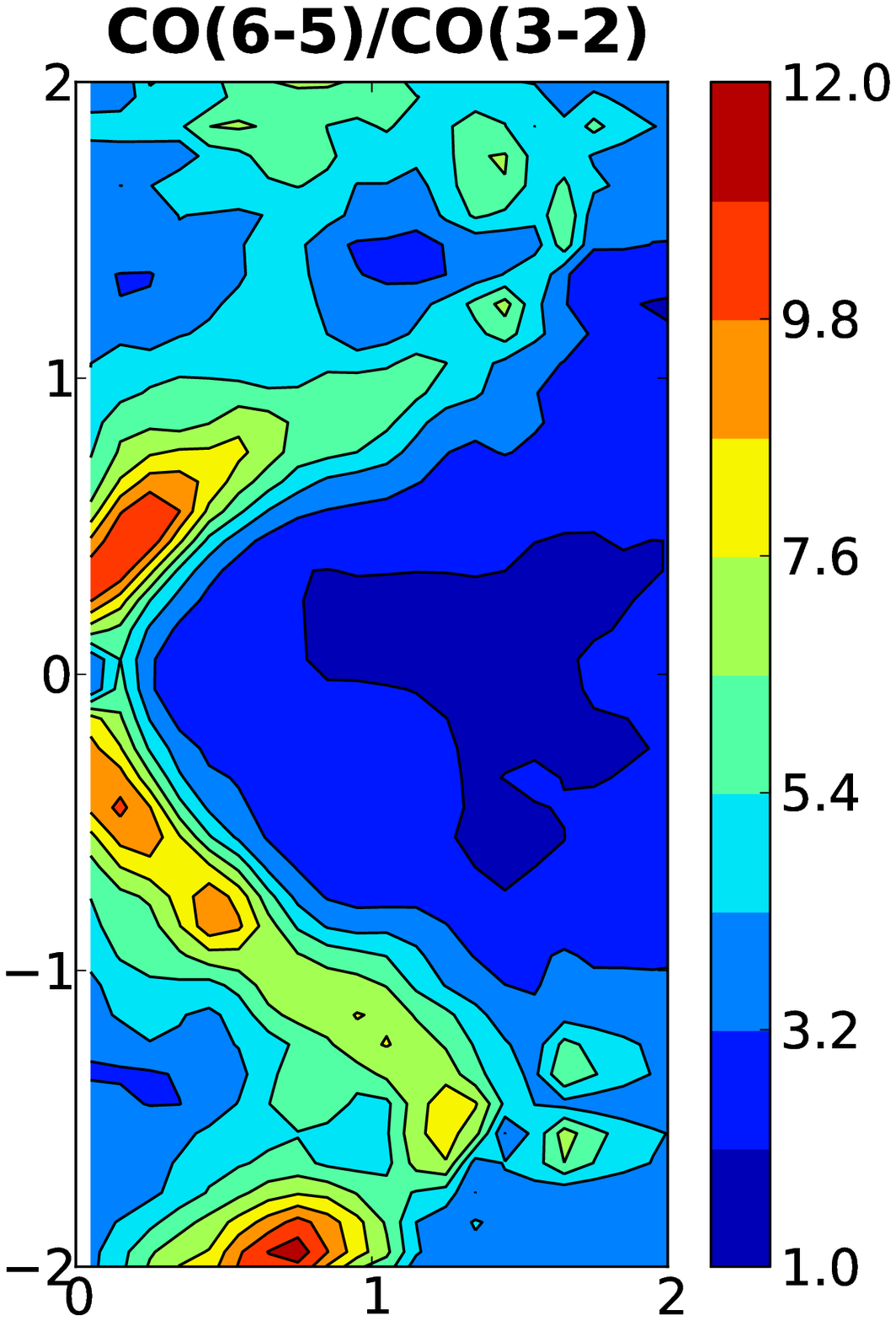}
\includegraphics[trim= 1.cm 0.cm .5cm 1.cm, clip, scale=0.34]{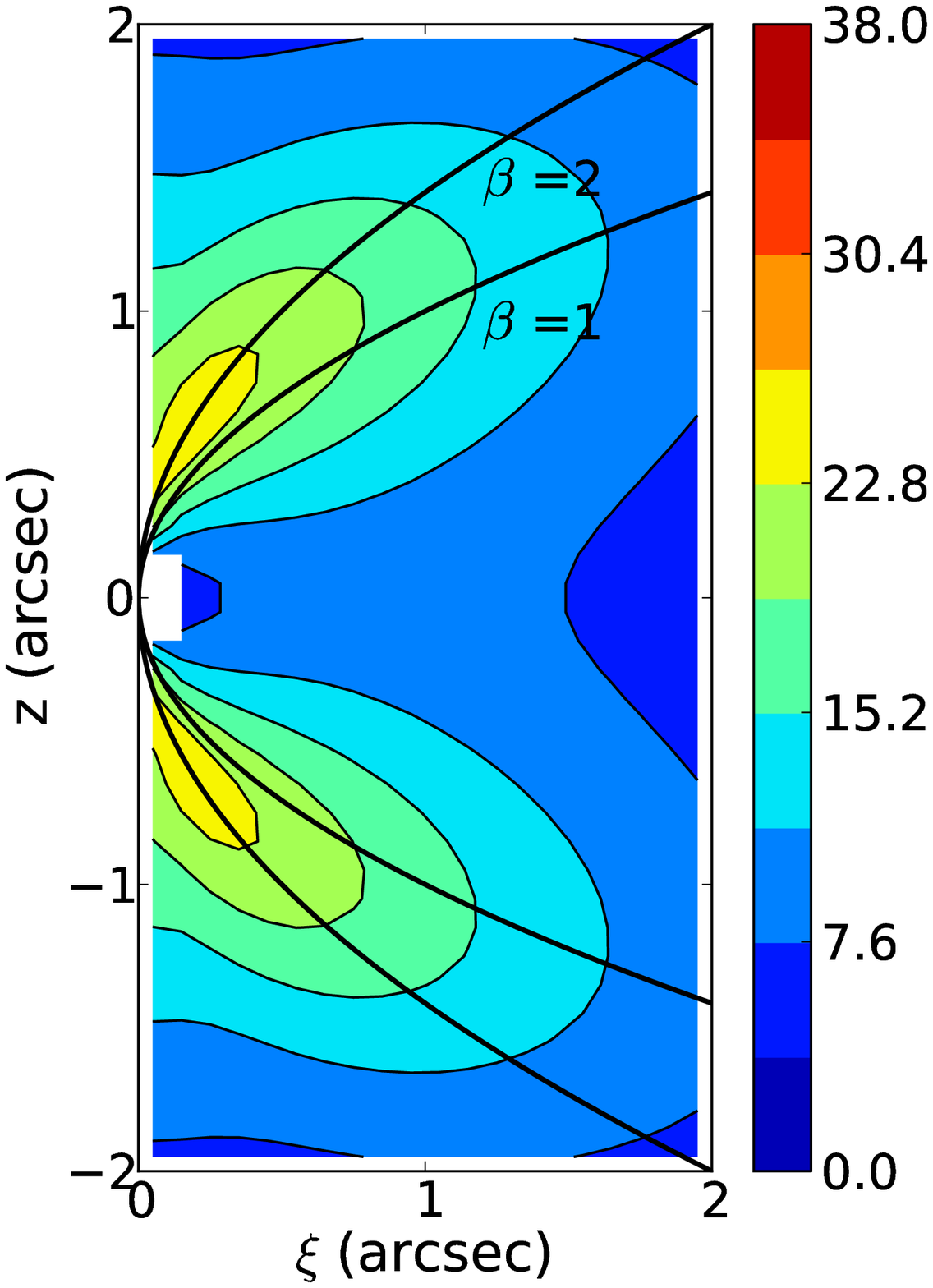} 
\includegraphics[trim= 1.5cm 0.cm .5cm 1.cm, clip, scale=0.34]{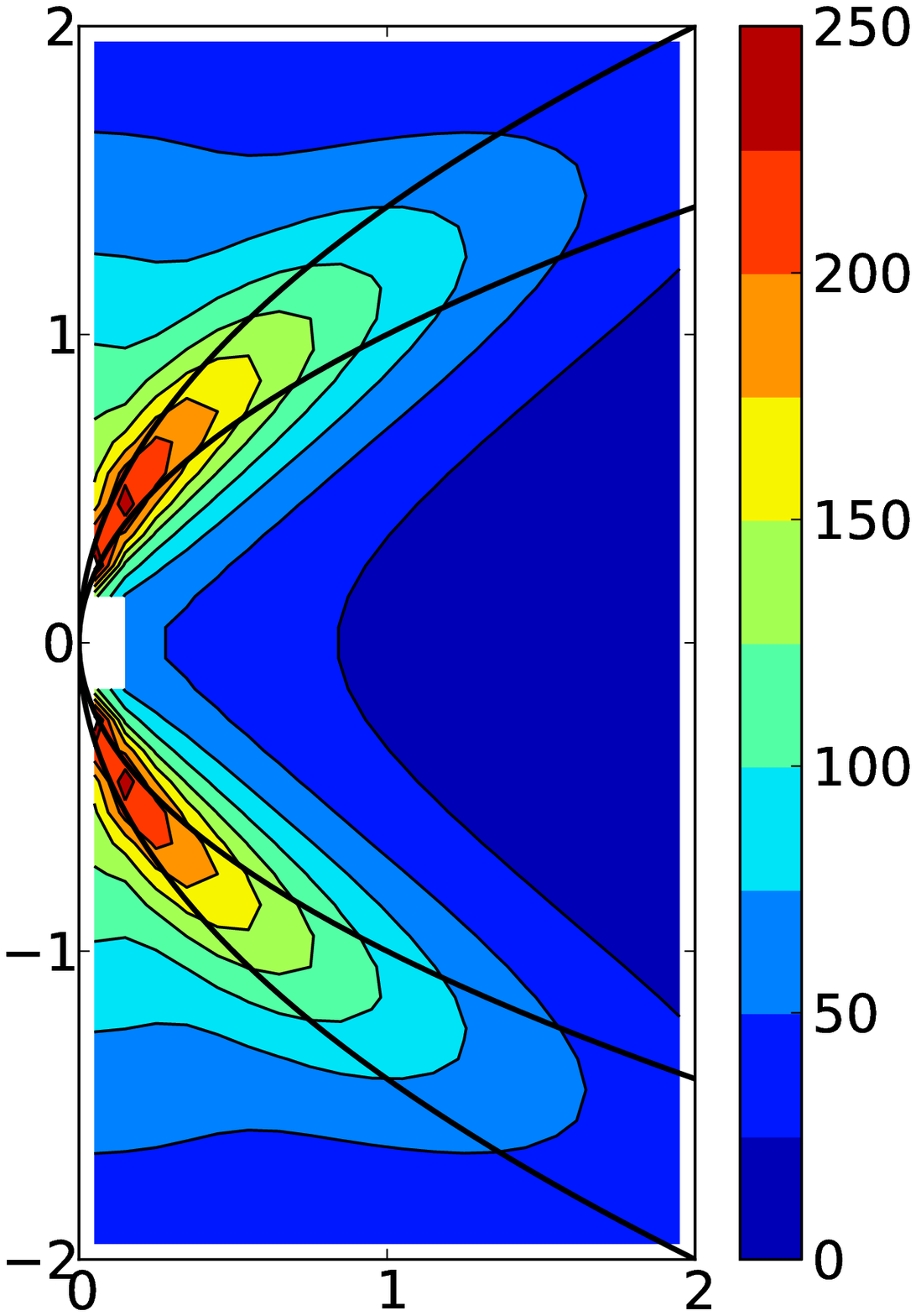}
\includegraphics[trim= 1.5cm 0cm .5cm 1.cm, clip, scale=0.34]{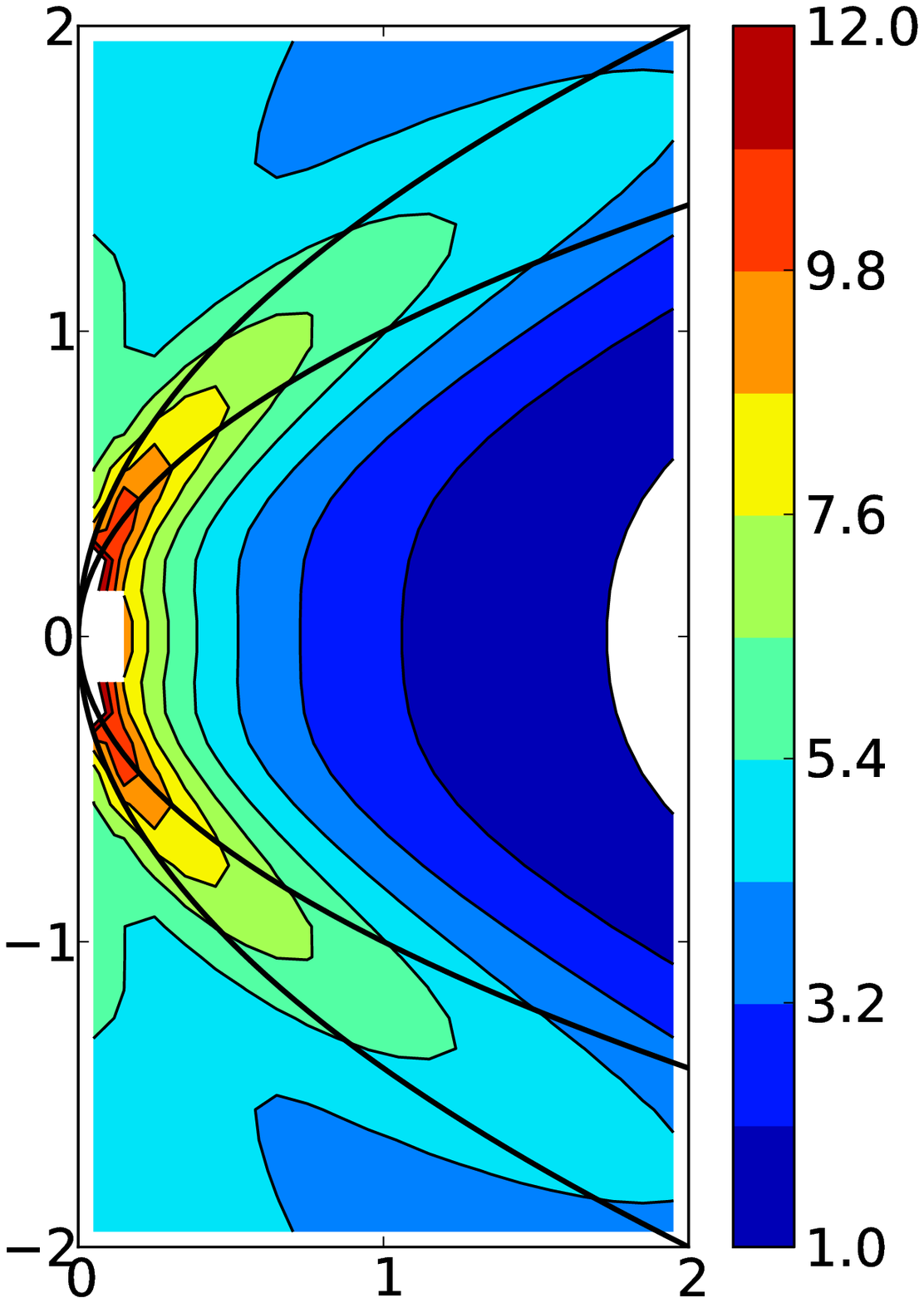}
\end{center}
\caption{Distribution of the effective densities multiplied by $r^2$ in the ($\xi$,$z$) meridian half-plane of the RR for \mbox{CO(3-2)} (left panel), \mbox{CO(6-5)} (middle panel) and the ratio \mbox{CO(6-5)}/\mbox{CO(3-2)} (right panel). The upper panels are for the solutions of the integral equation and the lower panels for the model described in the text. Parabolas corresponding to $\beta$=1 and $\beta$=2 are shown in the lower panels.}\label{Fig6}
\end{figure*}
 The present section deals with the first of these three steps. The integral equation is solved by iteration, using as input the fluxes, integrated over Doppler velocities, measured in each of the 1526 pixels that have been retained. Integration along the line of sight is made in steps of $x$ of 0.02$''$ with the space radius $r$ limited to the interval [0.1$''$, 3.5$''$]. The effective densities are defined on an array of 35 $\xi$ bins and 50 $z$ bins, each 0.1$''$ wide covering half a meridian plane of the star. The iterative process is stopped when the value of $\chi^2$ used to measure the quality of the agreement between observations and fluxes obtained by integration of the effective densities along the line of sight reaches a constant value. Various reasonable estimates of the measurement uncertainties have been tried in the definition of $\chi^2$, all giving very similar results. The uncertainties retained here are the quadratic sum of an absolute and a relative contributions, the former being 0.16 Jy$\times$arcsec$^{-2}$ for \mbox{CO(3-2)} and 2.2 Jy$\times$arcsec$^{-2}$ for \mbox{CO(6-5)} and the latter 20\% for each. Ten iterations are amply sufficient to achieve convergence. In order to illustrate the quality of the numerical resolution of the integral equation, we compare in Figure \ref{Fig5} the observed flux densities with those obtained by integration of the effective densities along the line of sight. To do so in a manageable way, we have grouped the data in 8 bins of $R$ and 8 bins of the position angle measured from the star axis, $\alpha=tan^{-1}(|y|/|z|)$. Figure \ref{Fig6} displays the distribution of the effective densities multiplied by $r^2$ in the meridian half-plane for \mbox{CO(3-2)}, \mbox{CO(6-5)} and their ratio. As the integral equation does not mix different values of $z$, it consists in fact of 50 independent integral equations, one for each $z$ bin, preventing a reliable evaluation of the effective density at large values of $|z|$, where input observations are scarce. For this reason, Figure \ref{Fig6} restricts $|z|$ and $\xi$ to the \mbox{[0, 2$''$]} interval.

 The observed morphology suggests introducing a parameter $\beta$ having the dimension of an angular distance, defined as $\beta=z^2/\xi$. Constant values of $\beta$ define parabolas having their axis in the equatorial plane and their summit at the star position. The region of large \mbox{CO(6-5)} to \mbox{CO(3-2)} ratio visible in the right panel of Figure \ref{Fig6} corresponds approximately to $\beta$$\sim$1 to 2. At larger distances from the star, parabolic arcs are seen in the form of ``wine glasses'' (Cohen M. et al. 2004), with axes along the star axis; they have no relation with the parabolas defined here, which are suited to the description of the gas envelope close to the equatorial torus. Indeed, beyond 2$''$ or so, the optimal description of the gas morphology has to evolve from an equator-dominated to a bipolar-dominated picture.

\begin{figure}[htb]
\begin{center}
\includegraphics[scale=0.4]{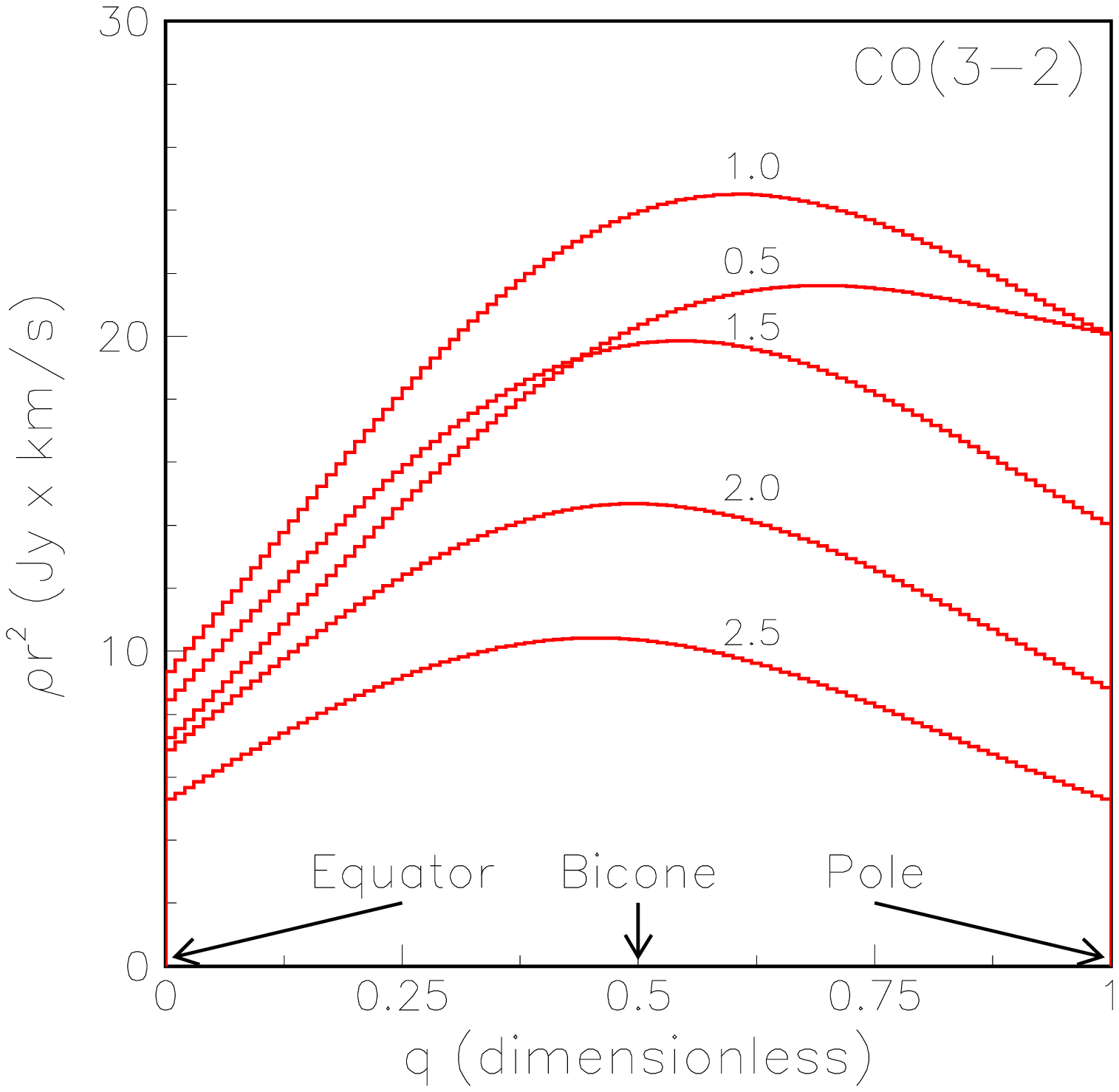}
\includegraphics[scale=0.4]{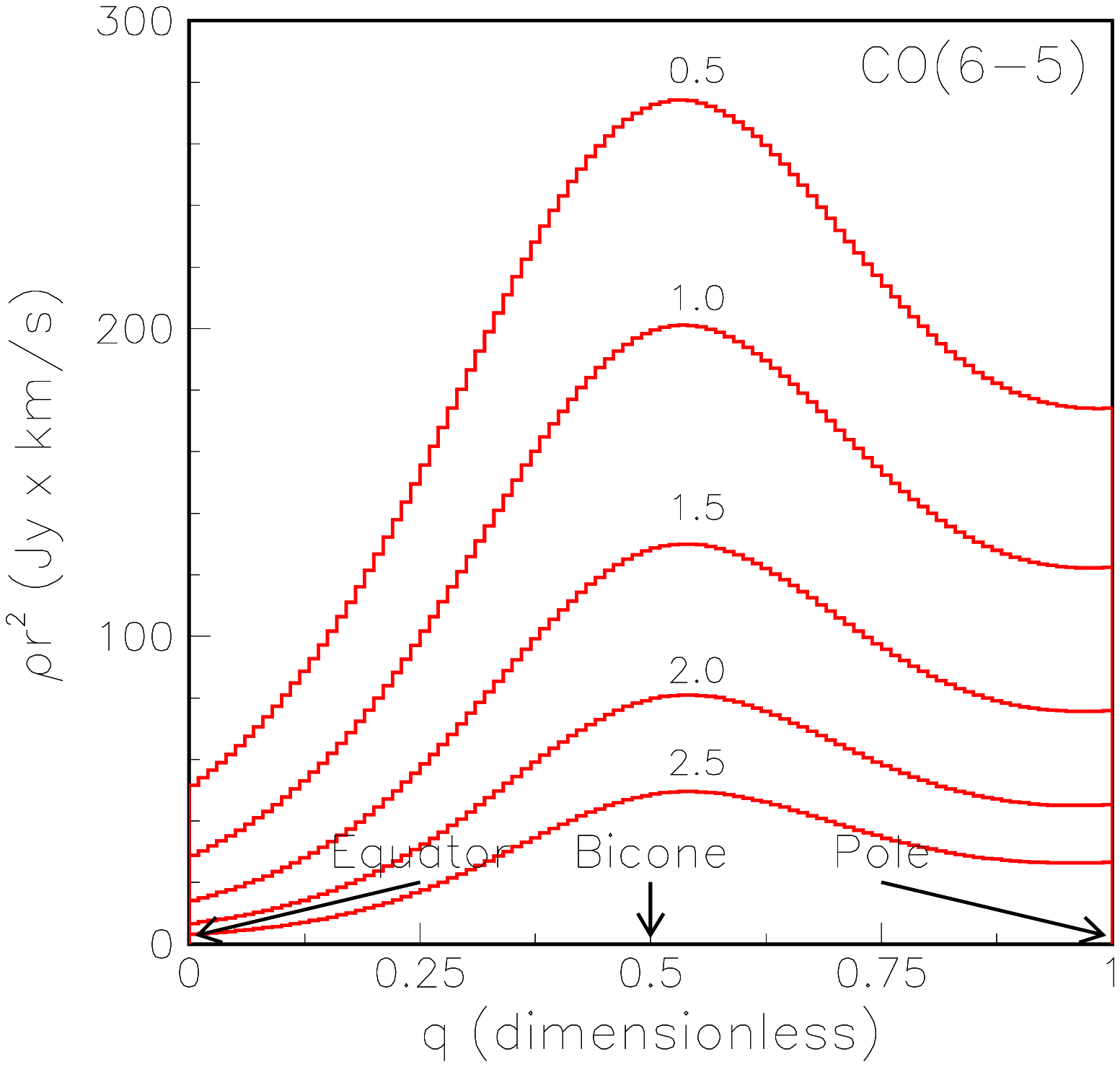}
\caption{Dependence on $q$=$1-exp(-\beta ln2/\beta_0)$ of the parameterized effective densities multiplied by $r^2$ for $r$=0.5$''$, 1.0$''$, 1.5$''$, 2.0$''$ and 2.5$''$ for \mbox{CO(3-2)} (left panel) and \mbox{CO(6-5)} (right panel).}\label{Fig7}
\end{center}
\end{figure}

\begin{figure}[ht]
\begin{center}
\includegraphics[scale=0.4]{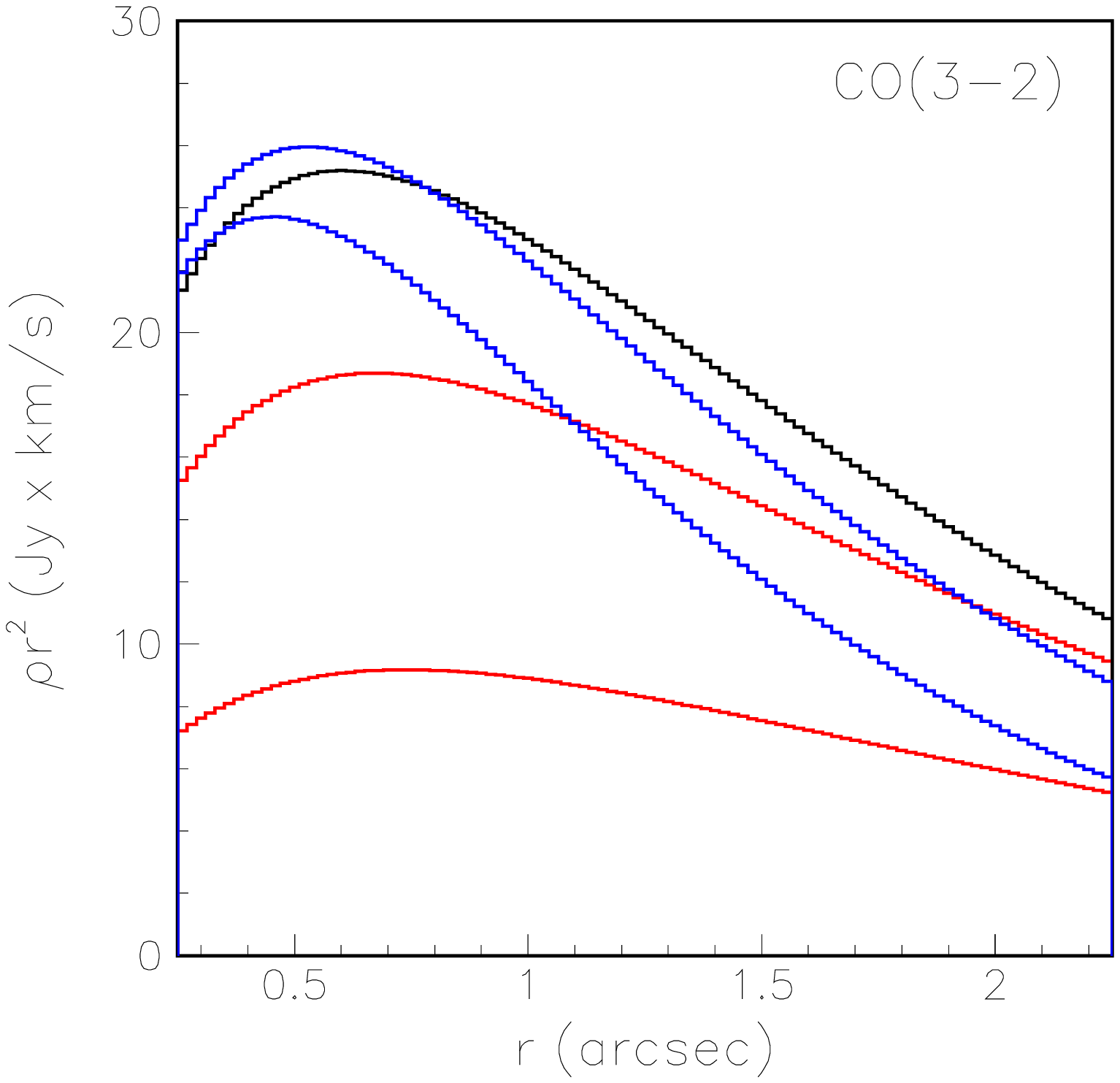}
\includegraphics[scale=0.4]{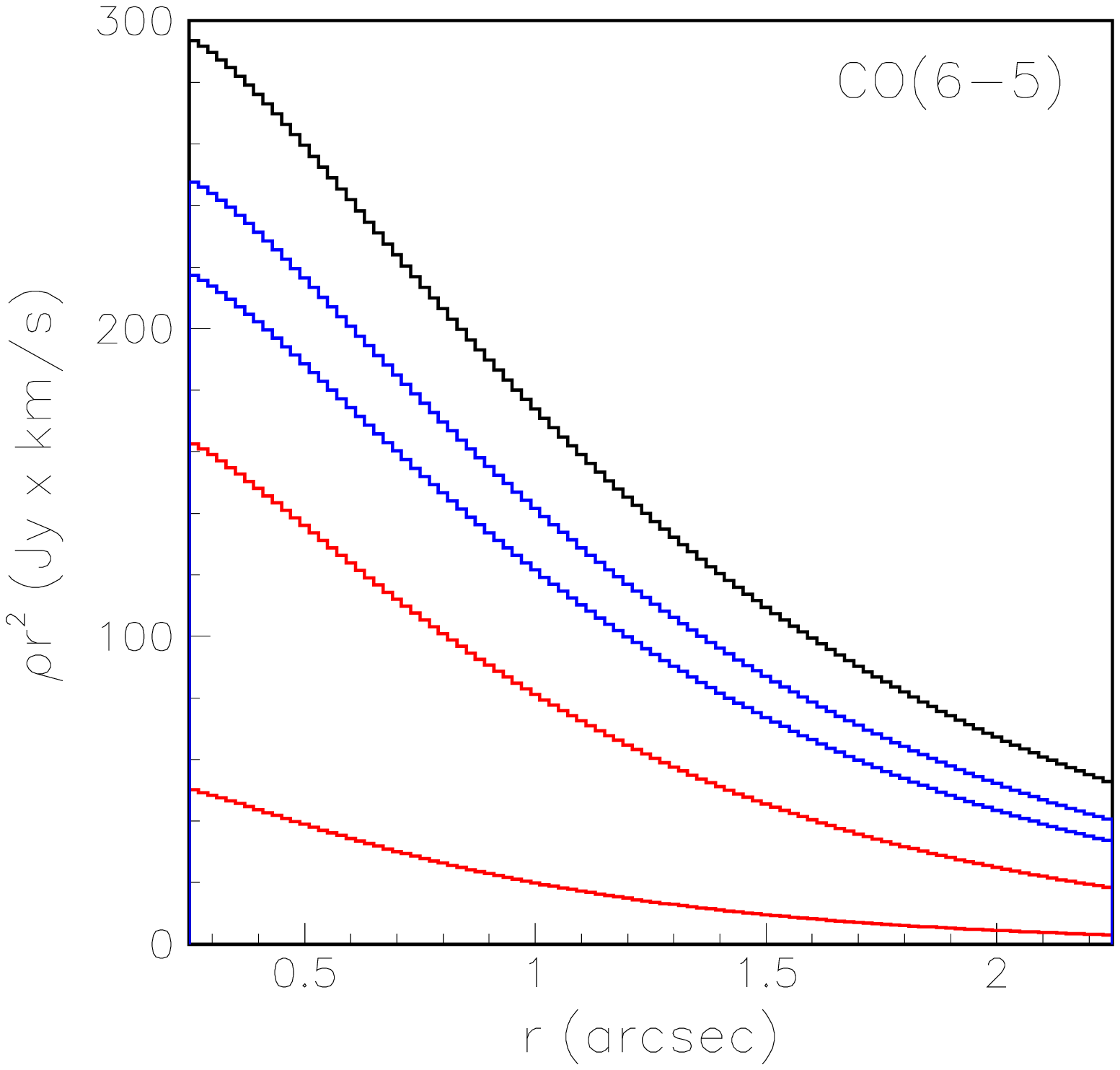}
\caption{Dependence on $r$ of the parameterized effective densities multiplied by $r^2$ for $q$ values of 0 and 0.25 (equator, red), 0.5 (bicone, black) and 0.75 and 1 (poles, blue) for \mbox{CO(3-2)} (left panel) and \mbox{CO(6-5)} (right panel).}\label{Fig8}
\end{center}
\end{figure}

 The present analysis does not require a parameterization of the effective densities, the following sections will use instead, for each data set, the array of numbers that has been obtained in the ($\xi$,$z$) meridian half-plane from the resolution of the integral equation. However, such a parameterization being useful to display the main features, we give a simple description of the effective densities in terms of the product of a function of $\beta$ by a function of $r$. As $\beta$ varies from zero at the equator to infinity at the pole, we use as variable the quantity $q=1-exp(-\beta ln2/\beta_0)$, which varies smoothly from zero at the equator to 1 at the pole, taking the value $\frac{1}{2}$ for $\beta=\beta_0$. Similarly, in order to describe the $r$-dependence, we use as variable the quantity $\psi=(r/r_0)^nexp(-r/r_0)$, which increases from zero at the origin to reach a maximum $n^ne^{-n}$ at $r=nr_0$ and then decreases exponentially to zero. Both $r_0$ and $\rho$ are given a dependence on $q$ in the form of a sum of a linear function and a Gaussian centred at 0.5 with a width $\sigma_0$. In summary, using the labels $eq$ for the equator, $bic$ for the bicone and $p$ for the poles,
\begin{subequations}
\label{eq3}
\begin{align}
&\rho=[\rho_{eq}+(\rho_p-\rho_{eq})q+\rho_{bic}g]\psi/r^2 \label{eq31} \\
&q=1-exp(-\beta ln2/\beta_0) \label{eq32} \\
&\psi=(r/r_0)^n exp(-r/r_0)  \label{eq33} \\
&r_0=r_{eq}+(r_p-r_{eq})q+r_{bic}g \label{eq34} \\
&g=exp[-\frac{1}{2}(q-\frac{1}{2})^2/{\sigma_0}^2 ]  \label{eq35}
\end{align}
\end{subequations}
The parameters have been adjusted to minimize the $\chi^2$ describing the quality of the fit in the region \mbox{($\xi$$<$2$''$,$|z|<$2$''$)} to the effective densities obtained above. The best fit values are listed in Table 1. We did not seek a parameterization giving a precise description of the effective densities in all details. Our ambition was only to give a description accounting for the main features and their most significant characteristics. Figure \ref{Fig6} (lower panels) displays the distribution of the parameterized effective densities multiplied by $r^2$ in the ($\xi$,$z$) meridian half-plane for \mbox{CO(3-2)}, \mbox{CO(6-5)} and their ratio, as was done in the upper panels for the effective densities themselves. Figure \ref{Fig7} displays the dependence on $\beta$ of the parameterized effective densities for different values of $r$ and Figure \ref{Fig8} that on $r$ for different values of $\beta$. They display broad latitudinal enhancements around the bicone, broader for \mbox{CO(3-2)} than for \mbox{CO(6-5)} and a steep decrease with radius. The enhancements are typically three times as wide as the beams and the difference between their appearances in \mbox{CO(3-2)} and \mbox{CO(6-5)} cannot be blamed on the different beam sizes used for the respective observations.
\begin{table*}
\caption{Best fit parameters to the \mbox{CO(3-2)} and \mbox{CO(6-5)} effective densities multiplied by $r^2$.}
\centering
\begin{tabular}{|c|c|c|c|c|c|c|c|c|c|}
\hline
 & $\beta_0$ & $n$ & $\sigma_0$ & $\rho_{eq}$ & $\rho_p$ & $\rho_{bic}$ & $r_{eq}$ & $r_p$ & $r_{bic}$ \\ 
\hline
\mbox{CO(3-2)} & 1.55 & 0.605 & 0.40 & 0.0655 & 0.534 & 0.593 & 1.21 & 0.737 & 0.025 \\ 
\hline
\mbox{CO(6-5)} & 1.36 & 0.215 & 0.207 & 0.334 & 1.26 & 1.27 & 0.60 & 0.84 & 0.19 \\ 
\hline
\end{tabular}
\end{table*}
\section{TEMPERATURE AND DENSITY DISTRIBUTIONS}
\label{sec:temperature}
 The preceding section studied the morphology of the CO envelope of the RR without paying particular attention to the absolute values taken by the effective density and the implication on the actual gas density and temperature. We address this issue in the present section.
\begin{figure*}[htb]
\begin{center}
\includegraphics[scale=0.32]{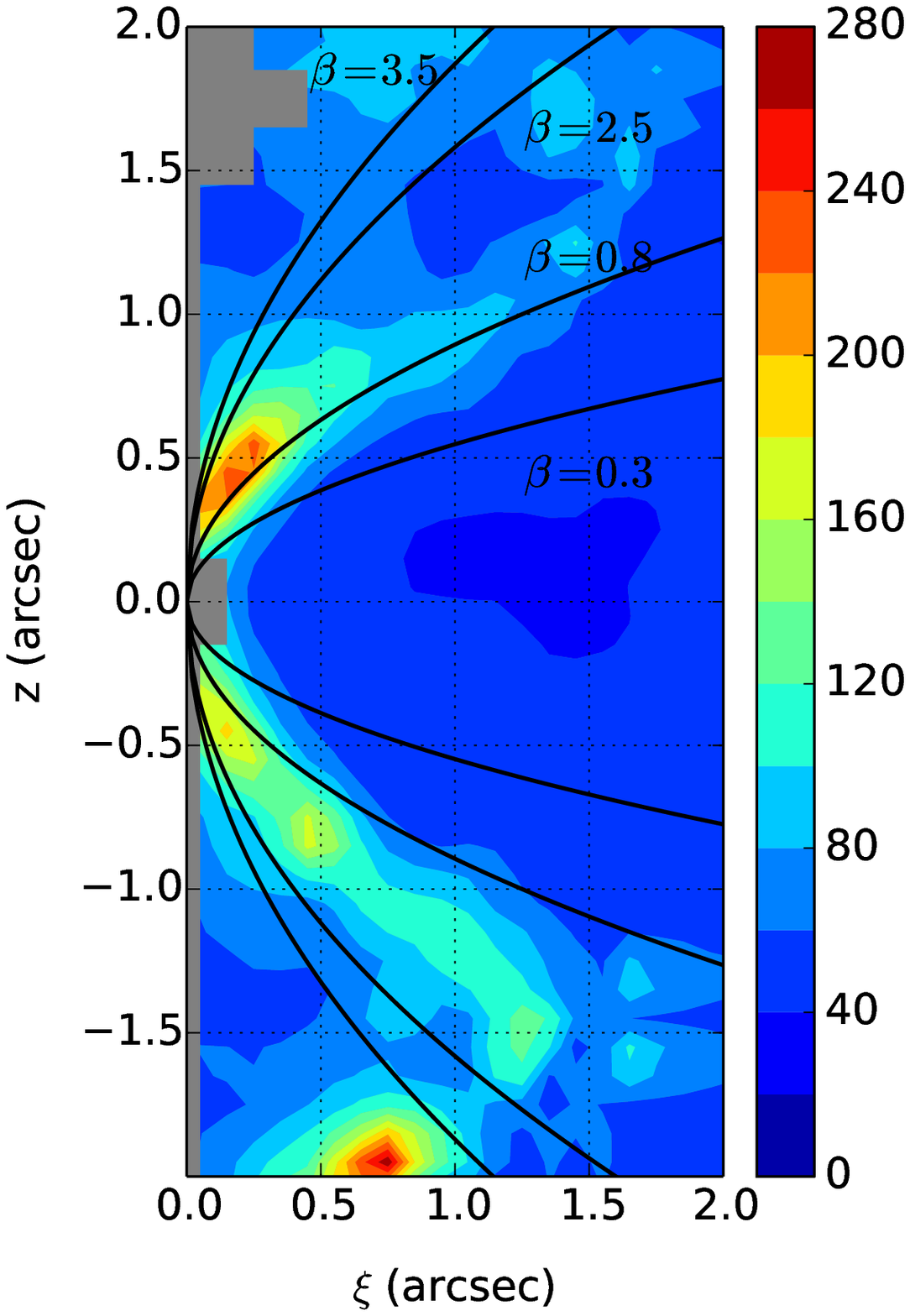}
\includegraphics[scale=0.44,width=6cm,height=6.8cm, trim=0.5cm -.2cm 0.5cm 0.,clip]{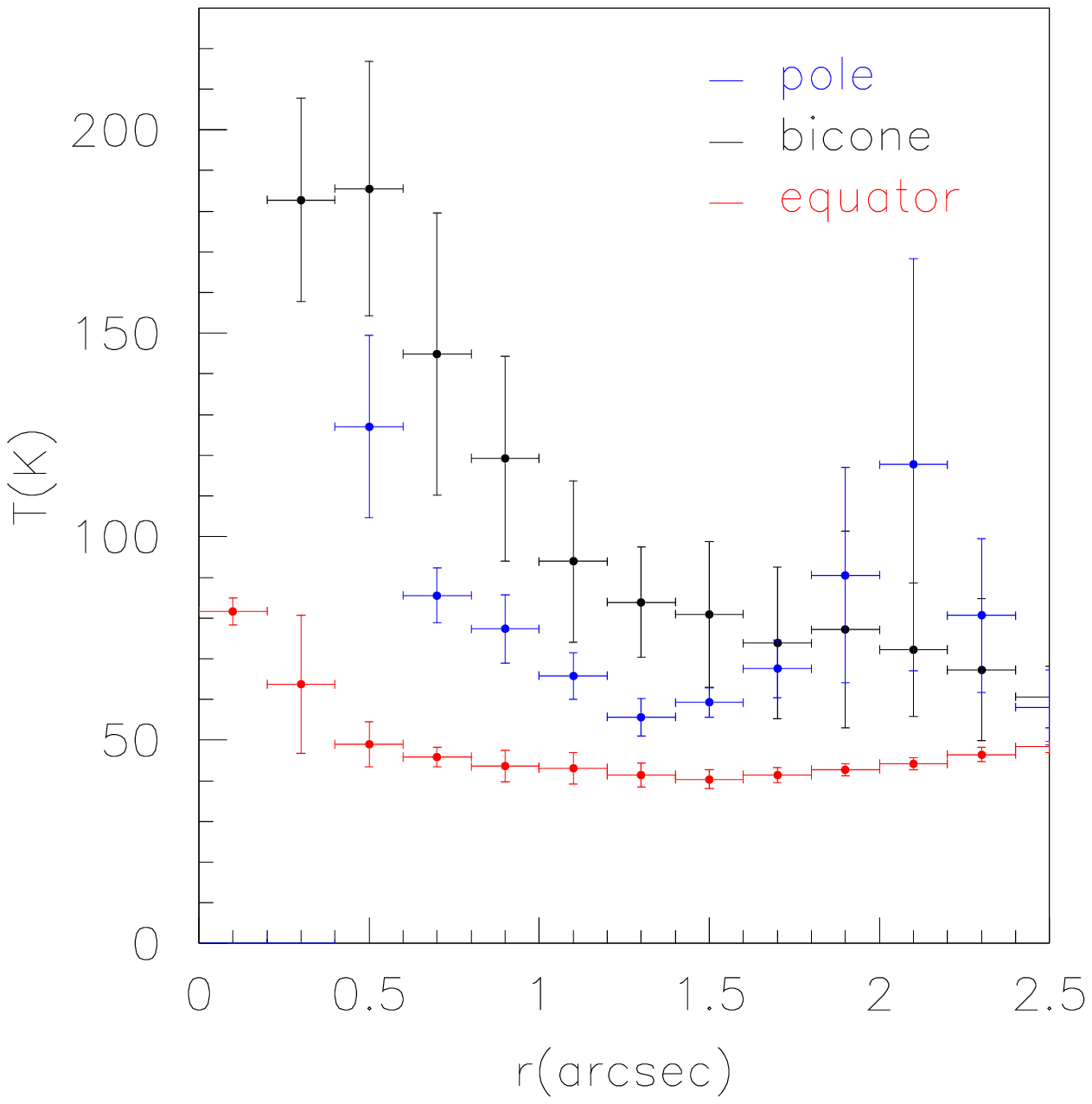}
\includegraphics[scale=0.32]{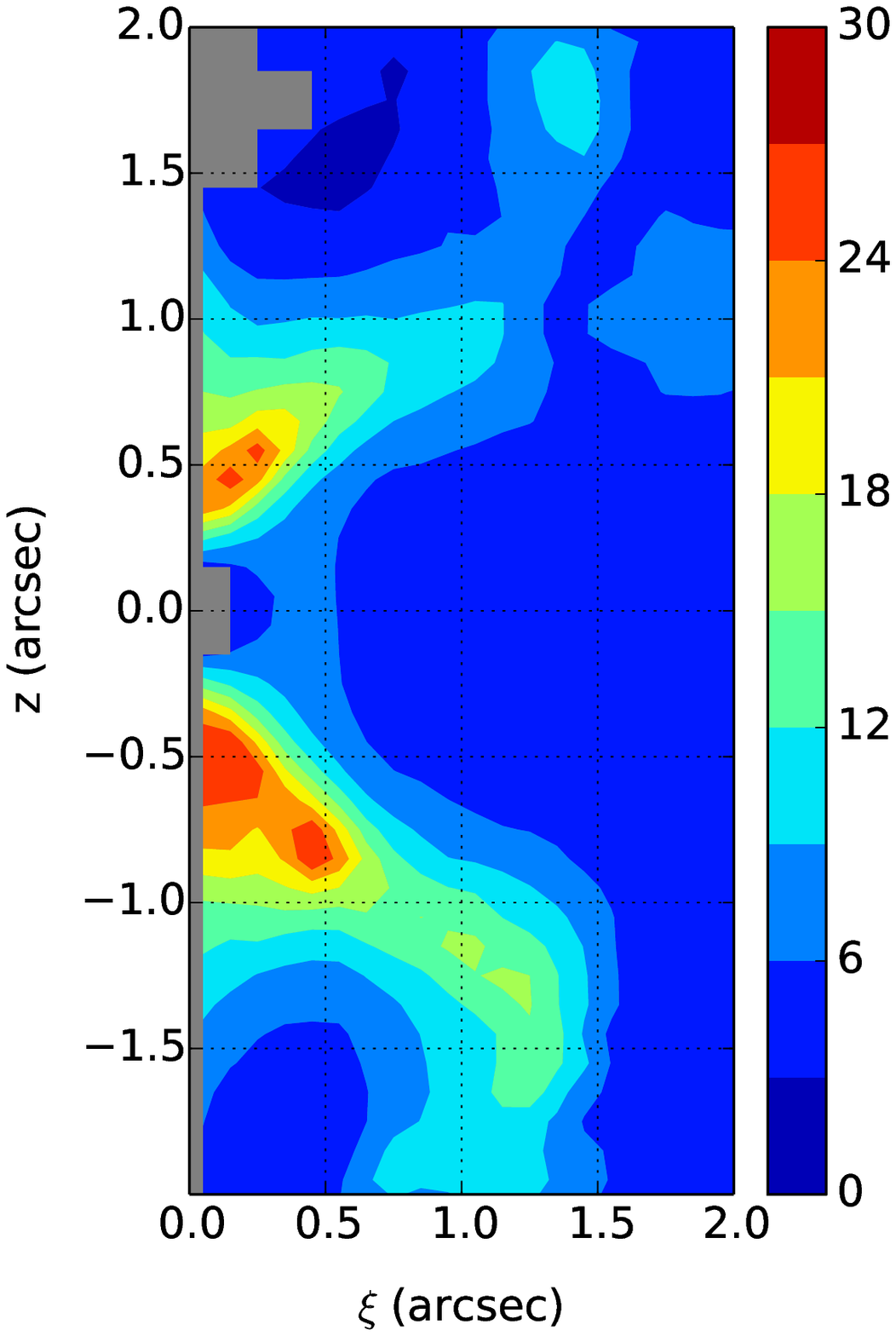}
\caption{Left: Map of temperatures in the half-meridian plane of the star obtained from the effective densities using Relation 4. The parabolas are for $\beta$=0.3, 0.8, 2.5 and 3.5 and define the sectors illustrated in the central panel. Middle: $r$-distribution of the gas temperature averaged over the three angular sectors delineated in the left panel: the red points are for the equatorial region (inside the $\beta=0.3$ parabola), the black points are on the bicone (between the $\beta=0.8$ and $\beta=2.5$ parabolas) and the blue points are for the polar region (outside the $\beta=3.5$ parabola). Error bars show the dispersion within each $r$ bin. Right: CO density (in molecules per cm$^3$) multiplied by $r^2$ (in arcsec$^2$).}\label{Fig9}
\end{center}
\end{figure*}
In a regime of thermal equilibrium, and in the absence of significant absorption, the effective density is the product of three factors: the actual gas density, which is the same for \mbox{CO(3-2)} and \mbox{CO(6-5)}; the population of the emitting rotational states (with angular momenta $J$=3 and 6 respectively); the emission probability. To a good approximation, the product of the second and third factors depends on temperature $T$ as $exp(-E_J/k_BT)/T$ with $E_J$ the energy of the emitting state and $k_B$ the Boltzmann constant. Therefore, the ratio $R_T$ of the effective densities obtained in the preceding section for \mbox{CO(6-5)} and \mbox{CO(3-2)} separately obeys the relation $R_Texp(-E_3/k_BT)$=$Cexp(-E_6/k_BT)$, with $C$ a known constant depending only on the values of $J$, 6 and 3 respectively. Hence \mbox{$ln(R_T/C)$=$E_3/k_BT-E_6/k_BT$} and 
\begin{equation}{\label{eq4}} 
k_BT=(E_6-E_3)/ln(C/R_T)
\end{equation}

In principle, Relation 4 provides a means to evaluate the gas temperature at any point in space once the value of $R_T$ is known. In practice, however, $dT/T$ being proportional to $dR_T/[R_Tln(R_T/C)]$, namely to $TdR_T/R_T$, the uncertainty attached to the measurement of $T$ increases as $T^2$, making higher temperatures increasingly difficult to evaluate. In particular, when $R_T$ approaches $C$, $T$ diverges, while when $R_T$ approaches 0, $T$ cancels. In the present case, $(E_6-E_3)/k_B$=82.5 K and $C$=15.6. The map of temperatures evaluated in this manner in the meridian half-plane of the RR is displayed in Figure \ref{Fig9}. While the temperature of the equatorial region decreases slowly from $\sim$60 K at $r$=0.3$''$ to $\sim$50 K at $r$=2$''$, the temperature of the outflow takes much higher values at short distance to the star, typically 200 K at $r$=0.3$''$, and decreases steeply with distance to reach some 80 K at $r$=1.5$''$. In comparison, the dust temperature quoted by Men'shchikov et al. (2002) decreases from $\sim$80 K at $r$=1$''$ to $\sim$60 K at $r$=2.5$''$. A hot spot at $z$$\sim$$-$2$''$ reaches a value of $\sim$270 K.

 The CO density, multiplied by $r^2$, is displayed in the right panel of Figure \ref{Fig9}. It is in good agreement with a model (Men'shchikov et al. 2002) assuming a dust over gas ratio of 1\% and a CO to H ratio of 2 10$^{-3}$.
\begin{figure}[hbt]
\begin{center}
\includegraphics[scale=0.4,trim=7cm 0. 4cm 0.,clip]{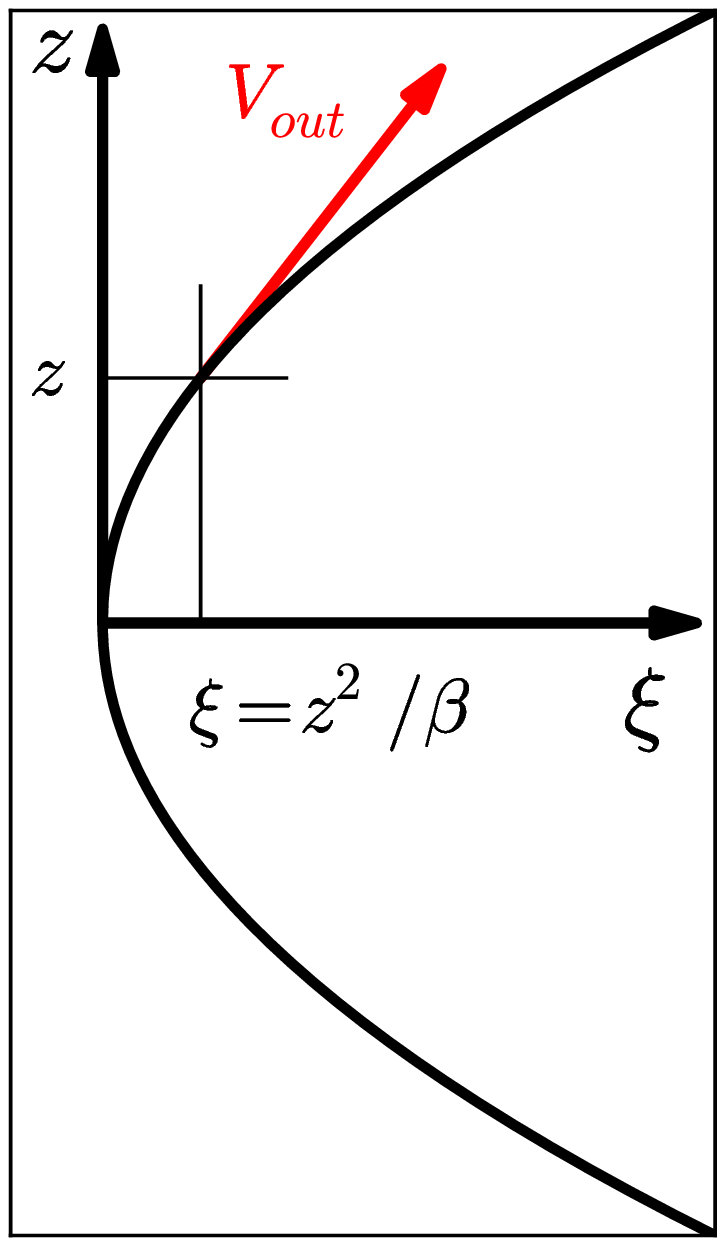}
\includegraphics[scale=0.4,trim=4cm 0. 5cm 0.,clip]{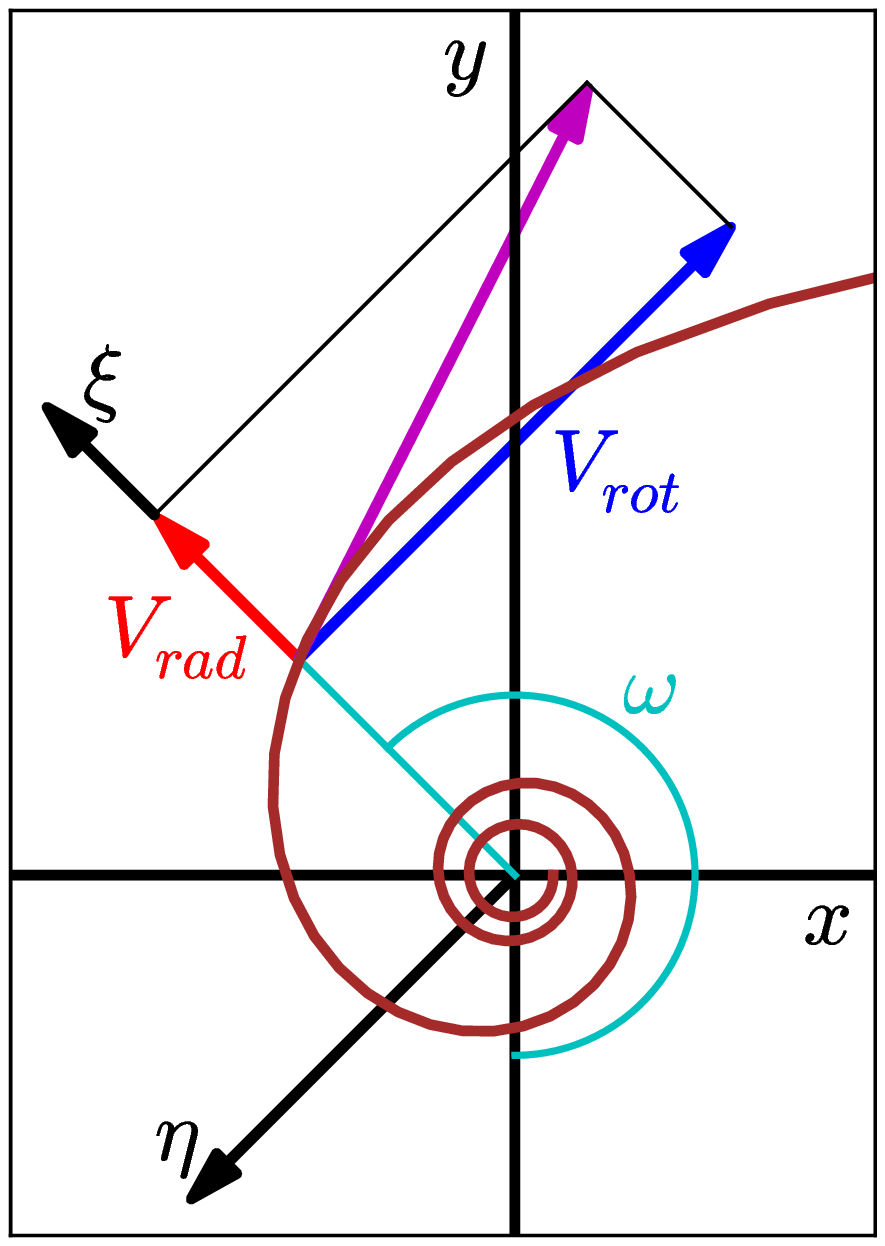}
\caption{Left: In the polar region ($\beta>\beta_0$) the gas velocity $V_{out}$ is confined to meridian planes ($\xi$,$z$) and tangent to parabolas of equation $z^2$= $\beta\xi$. Right: In the equatorial region ($\beta<\beta_0$) the gas velocity is confined to planes parallel to the equatorial plane and tangent to hyperbolic spirals with a constant radial component $V_{rad}$ and a rotation velocity $V_{rot}$ proportional to $r^{-k}$.}\label{Fig10}
\end{center}
\end{figure}
\begin{table*}
\caption{Best fit parameters $P$ of the joint fit to the \mbox{CO(3-2)} and \mbox{CO(6-5)} spectral maps. Also listed are the values of $\Delta^+$ and $\Delta$$^-$ measuring the sensitivity of the value of $\chi^2$ to small deviations of the parameter from its best fit value (see text).}
\centering
\begin{tabular}{|c|c|c|c|c|c|c|c|c|}
\hline
 & $\beta_0$ ($''$) & $V_{rot}$ (kms$^{-1}$) & $k$ & $V_{out}$ (kms$^{-1}$) & $V_{rad}$ (kms$^{-1}$) & $\theta$ ($^\circ$) & $\sigma_{p}$ (kms$^{-1}$) & $\sigma_{eq}$ (kms$^{-1}$) \\ 
\hline
P & 0.81 & 0.99 & 1.03 & 6.3 & 1.56 & 8 & 1.4 & 1.0 \\ 
\hline
$\Delta ^+$ & 0.15 & 0.15 & 0.10 & 0.8 & 0.26 & 8 & 0.7 & 0.3 \\ 
\hline
$\Delta ^-$ & 0.09 & 0.14 & 0.18 & 0.7 & 0.26 & 8 & 0.5 & 0.2 \\ 
\hline
\end{tabular}
\end{table*}

\section{GAS VELOCITY}
\label{sec:velocity}
 We use Relation (1) $ V_x$=$(x/\xi)V_{\xi}-(y/\xi)V_{\eta}+\theta V_z$ to evaluate the gas velocity components $V_{\xi}$ and $V_{\eta}$, the last term providing an evaluation of the small tilt $\theta$ once a value is assumed for $V_z$. We use the results of the preceding sections to compare the data with a simple model (Figure \ref{Fig10}) allowing for rotation about the star axis in the equatorial region and for a polar outflow, the parabolic separation between them being defined by a first parameter, $\beta_0$, expected to be of the order of 0.8$''$. In the equatorial region, we adopt as model $V_{\eta}$=$V_{rot} r^{-k}$, with $k$$\sim$0.5 for a Keplerian motion, and allow for a constant radial expansion, normal to the star axis, $V_{\xi}$=$V_{rad}$, $V_z$=0. In the polar region, on the contrary, we take $V_{\eta}$=0 and velocities in the meridian plane tangent to the parabolas associated with constant $\beta$ values, $V_{\xi}$=$2V_{out}z/\sqrt{\beta^2+4z^2}$ and $V_z$=$V_{out}\beta/\sqrt{\beta^2+4z^2}$. Here, $V_{out}$=$\sqrt{V_{\xi}^2+V_z^2}$. Moreover, we allow for a small tilt $\theta$ of the star axis with respect to the sky plane, producing a velocity $\theta V_z$ along the line of sight.

 As the gas velocities are the same for \mbox{CO(3-2)} and \mbox{CO(6-5)} emission, we fit both sets of data jointly, each set being given the effective densities obtained in the preceding section. The six model parameters are adjusted by minimizing the value of $\chi^2$ that measures the quality of the fit to the velocity spectra associated with each pixel of each of the two data sets. We require $R$ to be in the interval [0.2$''$, 2.5$''$] and $r$ in the interval \mbox{[0.2$''$, 3.5$''$]}. We find that the fits are significantly improved when allowing for some Gaussian smearing of the velocity distributions, the best values of their $\sigma$'s being $\sigma_{p}$=1.4 kms$^{-1}$ in the polar region and $\sigma_{eq}$=1.0 kms$^{-1}$ in the equatorial region. Such large values are likely to account for imperfections of the model rather than for a physical dispersion of the space velocities. Setting them to zero deteriorates the quality of the fit but does not alter the values taken by the other parameters. The best fit values of the adjusted model parameters are listed in Table 2 together with quantities $\Delta$$^+$ and $\Delta$$^-$ measuring the sensitivity of the value of $\chi^2$ to small deviations of the model parameters from their best fit values. They are defined such that when a parameter having best fit value $P$ varies in the interval [$P-$$\Delta$$^-$,$P+$$\Delta$$^+$], $\chi^2$ does not exceed its minimal value by more than 5\%.
\begin{figure*}[htb]
\begin{center}
\includegraphics[scale=0.44,trim=1cm 0. .5cm 0.,clip]{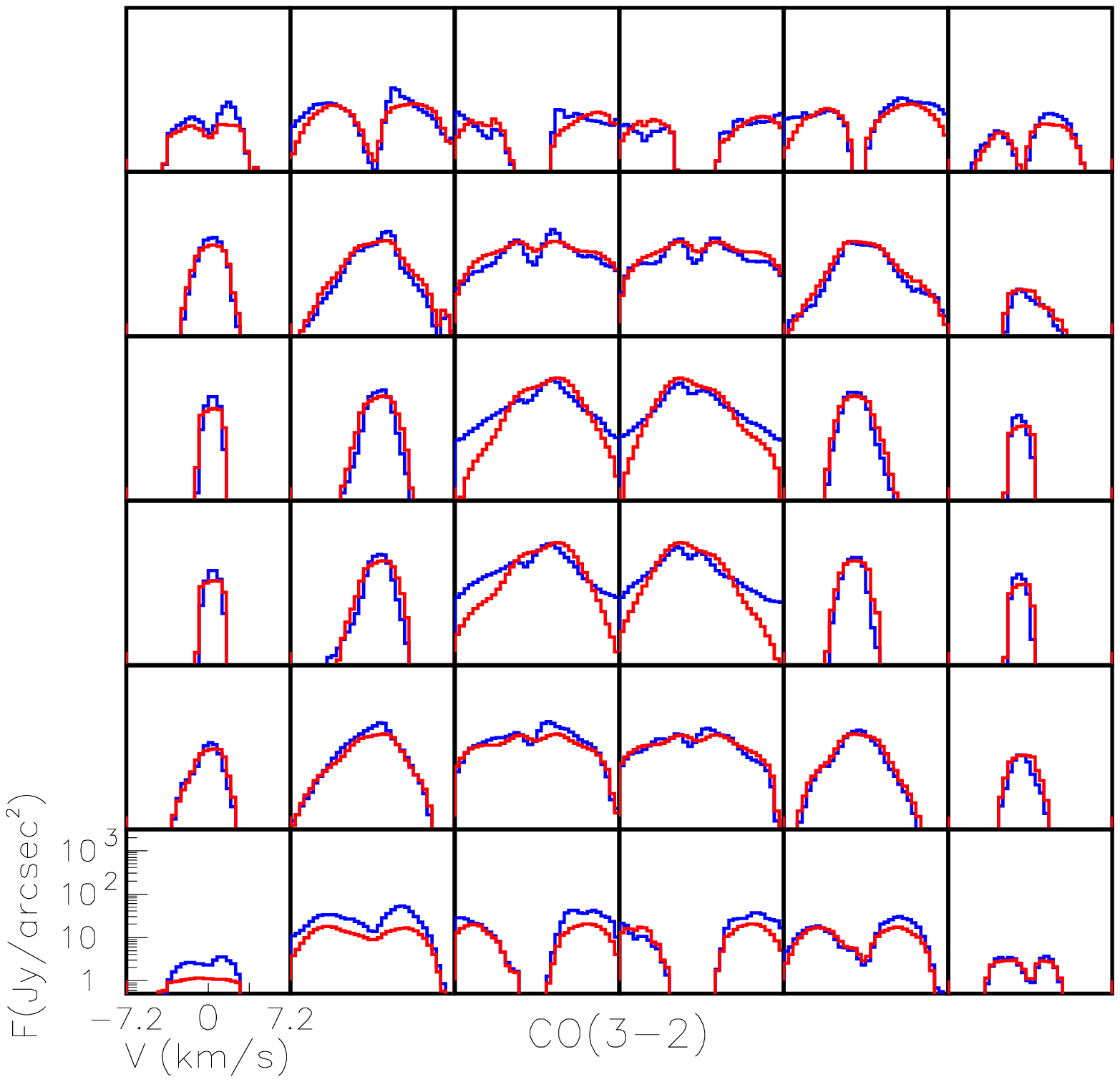}
\includegraphics[scale=0.44,trim=.5cm 0. 1.cm 0.,clip]{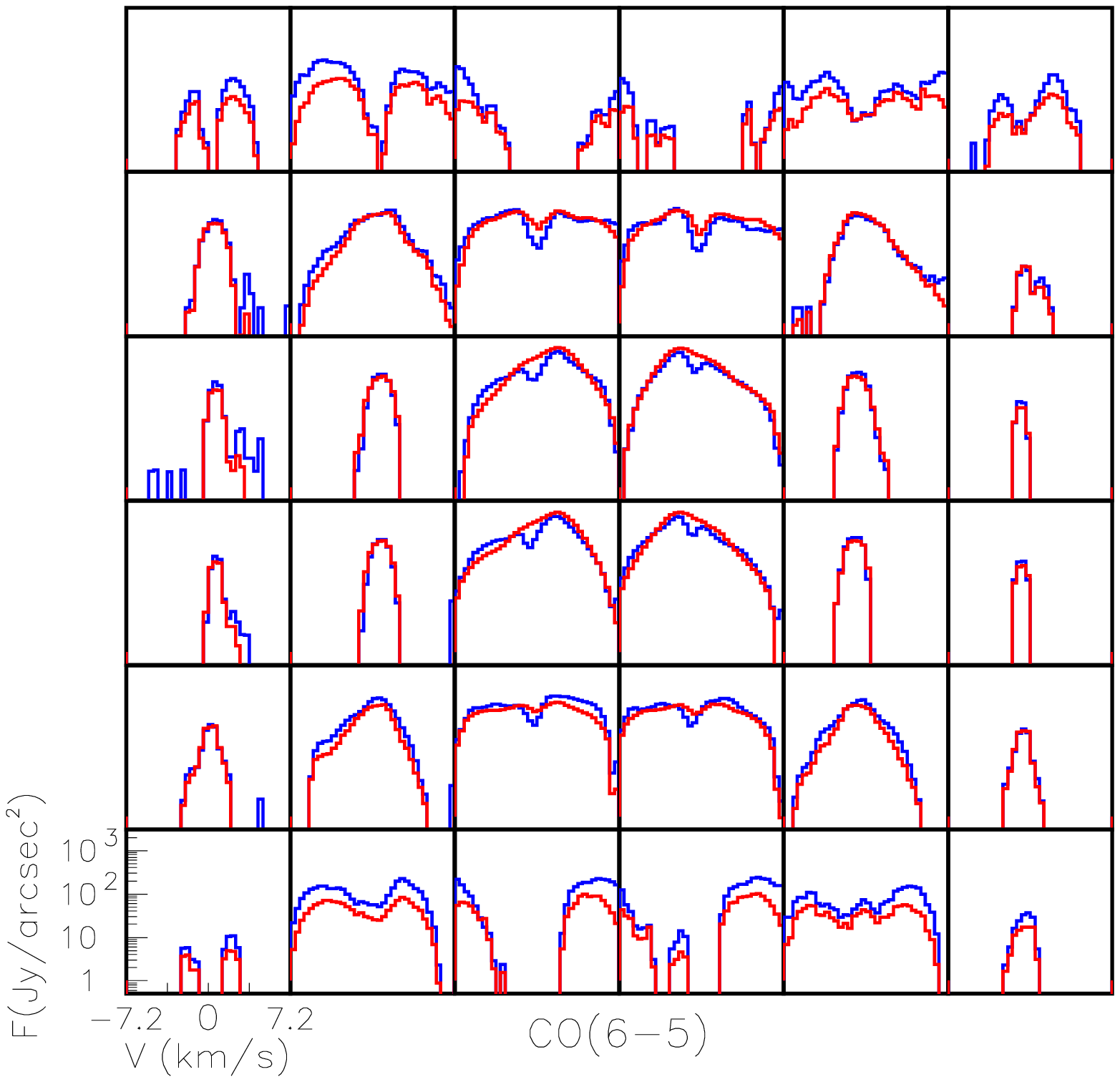}
\caption{{{\mbox{CO(3-2)} (left) and CO(6-5) (right) velocity spectra for data (blue) and model (red) averaged over groups of 49 pixels, each group covering 0.7$''$$\times$0.7$''$, the whole map covering 4.2$''$ $\times$4.2$''$}}.}\label{Fig11}
\end{center}
\end{figure*}
The quality of the fit is surprisingly good in view of the crudeness of the model. It is illustrated in Figure \ref{Fig11} for \mbox{CO(3-2)} and \mbox{CO(6-5)} separately. While each of the 1526 velocity spectra associated to the pixels that have been retained contributes separately to the value of $\chi^2$, they have been grouped for convenience into only 36 spectra in Figure \ref{Fig11}, each group covering a square of 7$\times$7=49 pixels.

 Allowing for additional parameters trivially improves the quality of the fit, but we have been unable to think of a simple specific addition that would do so very significantly. In particular, allowing for a transition region between the polar outflow and the equatorial rotating torus does not bring much improvement, the data being satisfied with a sharp transition. Similarly when allowing for variations of the parameters $V_{rot}$, $V_{out}$ and $V_{rad}$ across the regions where they operate. It is difficult to give precise evaluations of the uncertainties attached to the parameters: the quality of the data, the validity of the approximations made and the crudeness of the model prevent doing it reliably. However, a few general results can be safely stated:

 1) Two regions, equatorial and polar, coexist, hosting very different velocity fields. To a good approximation, they are separated in the meridian half-plane by a parabola of equation $z^2/\xi$=0.81$''$.

 2) The evidence for rotation of the equatorial torus, with velocity of $\sim$1 kms$^{-1}$ at $r$ =1$''$, is overwhelming and its $r$-dependence requires a power index $k$ of order unity. Both density and rotation velocity are homogeneously distributed across the torus rather than on a thin disk. The need for radial expansion, at the level of $\sim$1.6 kms$^{-1}$, arises from the clear separation between positive and negative velocities at small values of $y$. Lacking such a component implies the presence of a peak at very low velocities.

 3) The evidence for a polar outflow is equally overwhelming. The choice made here of a parabolic flow in the meridian plane cannot be claimed to be unique. However, it gives significantly better results than a radial outflow, in addition to being more sensible from a pure hydrodynamic point of view once a parabolic separation is adopted between torus and polar outflow. Moreover, it implies an $r$-dependence of the Doppler velocity that fits well the data, making the introduction of a velocity gradient unnecessary.

 4) With respect to the sky plane, the star axis is inclined by $\sim$8$^\circ$ from the North away the Earth.

\section{ASYMMETRIES}
\label{sec:asymmetries}
\begin{figure*}[htb]
\begin{center}
\includegraphics[scale=0.38,trim=1cm 0. 1.5cm 0.,clip]{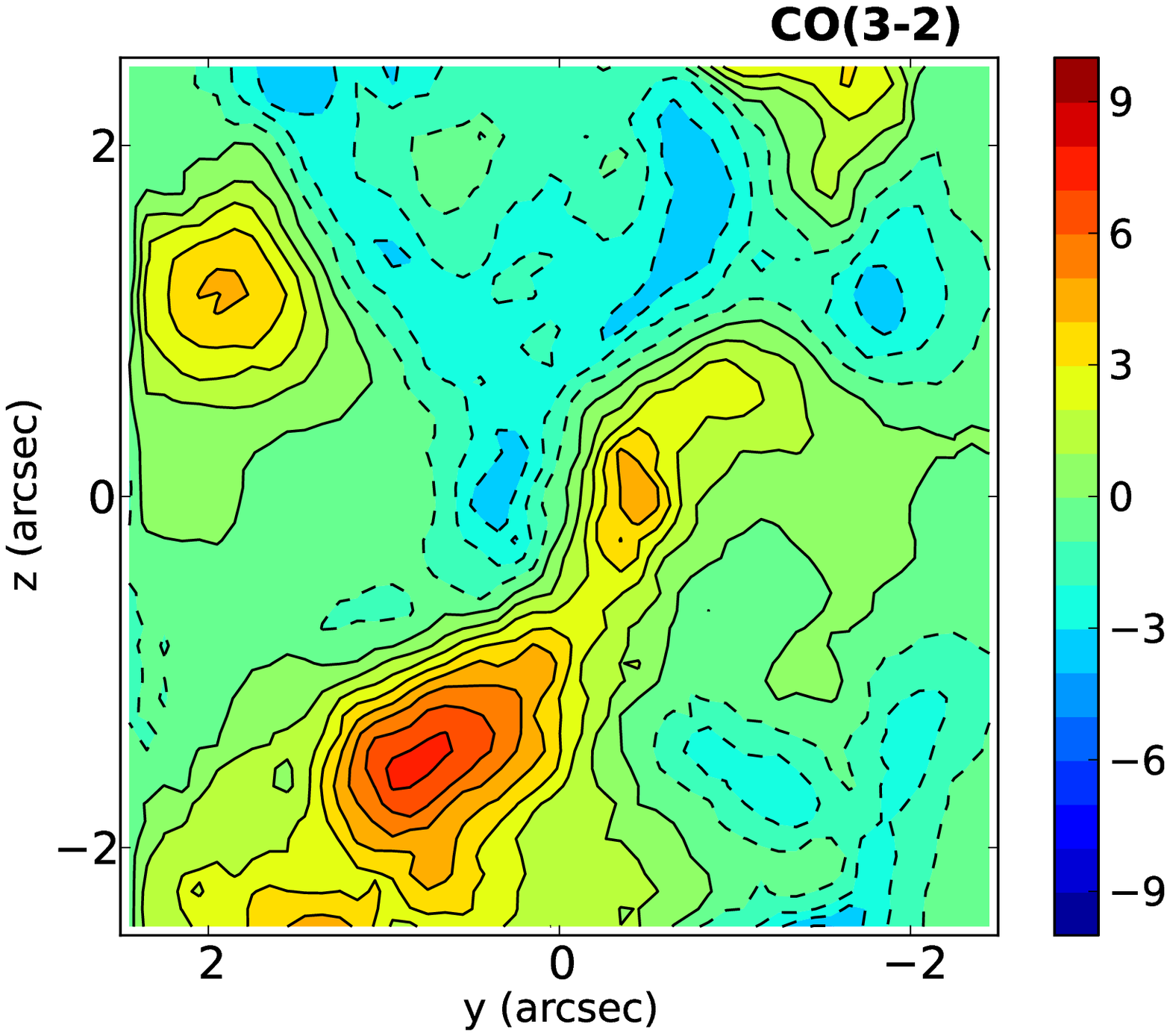}
\includegraphics[scale=0.38,trim=1cm 0. 1.5cm 0.,clip]{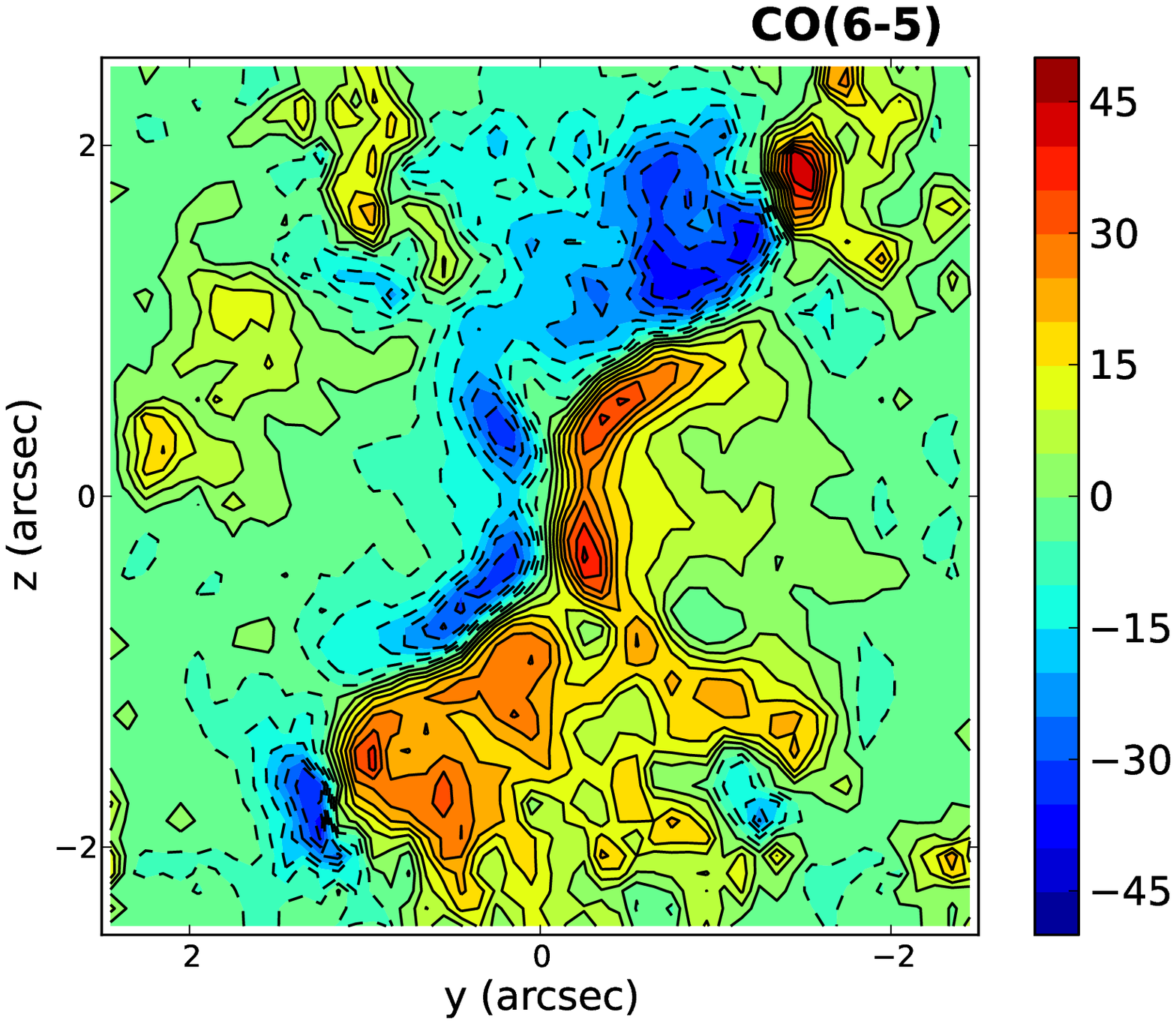}
\caption{Sky maps of the deviation from full symmetry (see text) multiplied by $R$ of the measured fluxes for \mbox{CO(3-2)} (left panel) and \mbox{CO(6-5)} (right  panel). Units are Jy$\times$kms$^{-1}$$\times$arcsec$^{-2}$. The south-eastern excess reaches $\sim$70\% of the symmetric value at maximum.}\label{Fig12}
\end{center}
\end{figure*}

 In the preceding sections, it was usually assumed that the gas effective density was invariant by rotation about the star axis and even, in several occasions, by symmetry with respect to the star equatorial plane. It was already remarked that these were crude approximations from which important deviations were present at large distances from the star. We map these in Figure \ref{Fig12} as deviations from $<F(y,z)>$=$\frac{1}{4}[F(y,z)+F(-y,z)+F(y,-z)+F(-y,-z)]$. More precisely, the quantity displayed is $[F(y,z)-<F(y,z)>]R$ where the factor $R$ gives a better balance between deviations at short distances and large distances from the star. Indeed, it would be the proper factor to be used if the effective densities would decrease with $R$ as $R^{-2}$ and if the uncertainties were purely statistical. The main feature is an important south-east excess, particularly enhanced in the region of the bicone where it reaches $\sim$70\%.
This observation addresses several questions: how reliable is it? which effect does it have on the results of the preceding sections? and which physics interpretation does it suggest?

 The presence of an excess on the eastern limb of the bicone, larger in the southern than in the northern region, can be safely asserted. We have checked its robustness by varying the parameters used in reducing the data, such as centring the sky maps or adjusting the baselines. Moreover, its presence in both the \mbox{CO(3-2)} and \mbox{CO(6-5)} data makes it unlikely that it could be blamed on imperfections of the calibration and/or imaging. However, to take seriously the finer details visible at large distances from the star, such as the fluctuations observed in the north-western part of the \mbox{CO(6-5)} map, would require reprocessing the images as done by Bujarrabal et al. (2013b).

The effect of such an excess $-$and other deviations from symmetry$-$ on the results of the preceding sections is simply to deteriorate the quality of the fits that have been performed but not to alter the results which have been stated in the approximation of rotation symmetry and/or north-south symmetry. As the magnitude of the excess is commensurate with that of the symmetric model, its contribution to $\chi^2$ is important and prevents attempting more detailed modelling than presented in the preceding sections without first accounting for the observed asymmetries.

 We are unable to state whether the observed excess is the result of temperature or of actual gas density, or both. A detailed study of its properties would probably allow for a reliable physics interpretation and provide very valuable information toward a better understanding of the Red Rectangle.

\section{CONTINUUM AND DUST}
\label{sec:continuum}
The space resolution and the sensitivity of the interferometer for the observation of the continuum are not sufficient to allow for a detailed study of the dust morphology. The projections of the flux density distributions on the $y$ and $z$ axes, displayed in Figure \ref{Fig2}, are well described by Gaussians having the following $\sigma$ values: at 345 GHz, $\sigma_y$=0.26$''$ and $\sigma_z$=0.23$''$ and at 690 GHz, $\sigma_y$=0.22$''$ and $\sigma_z$=0.13$''$. The ratio $\sigma_z/\sigma_y$ takes values of 0.88 and 0.60 at 345 GHz and 690 GHz respectively, revealing an elongation of the dust along the $y$ axis, at variance with the distribution of the line, which reveals an elongation of the gas along the $z$ axis. This is consistent with dust being concentrated in the equatorial region. The values of $\sigma_y$ and $\sigma_z$ are dominated by the beam size, twice as large at 345 GHz as at 690 GHz, consistent with a source concentrated at short distance to the star (Men'shchikov et al. 2002). The integrated fluxes are 0.66$\pm$0.10 Jy at 345 GHz and 4.0$\pm$0.6 Jy at 690 GHz, with a 690 GHz to 345 GHz ratio of 6.0 compared with a ratio of $\sim$5 corresponding to the measured SED (see Men'shchikov et al. 2002).

\section{SUMMARY AND CONCLUSIONS}
\label{sec:conclusions}
 ALMA observations of the CO emission of the Red Rectangle, of an unprecedented quality, have been analysed with the aim to reveal the main features of the morphology and kinematics of the gas envelope. The analysis was performed in a spirit of simplicity, with limited ambitions in terms of precision and sensitivity: it used the image processing provided by the ALMA staff while the reprocessing performed by Bujarrabal et al. (2013b) should allow for the exploration of finer details than was possible in the present work. 

The effective density, combining actual density and temperature, was reconstructed in space under the hypothesis of rotation symmetry about the star axis and assuming that absorption is negligible. Both are very crude approximations that limit the scope of the study. In principle, $^{13}$CO observations available in the ALMA data set should shed light on the validity of the optically thin approximation; however, the weakness of the line would require reprocessing the image in order to obtain reliable results. The effective density was observed to decrease with distance faster than $r^{-2}$, requiring an additional exponential factor with characteristic length at the arcsecond level, and to vary smoothly as a function of star latitude, reaching a maximum at latitudes between 45$^\circ$ and 60$^\circ$ typically. Comparison between the \mbox{CO(3-2)} and \mbox{CO(6-5)} effective densities provides an evaluation of the gas temperature, observed to decrease slowly with distance in the equatorial region, from $\sim$60 K at 0.3$''$ to $\sim$50 K at 2$''$. In the polar region, the temperature takes much higher values at short distance to the star, typically 200 K at 0.3$''$, and decreases steeply with distance to reach some 80 K at 1.5$''$. A crude model of the calculated effective densities has been presented.

 The study of the gas kinematics has revealed a sharp separation between the equatorial and polar regions. To a good approximation, the former is a parabolic torus in rotation about the star axis and expansion outward from it. The rotation velocity is of the order of 1 kms$^{-1}$ at a distance of 1$''$ and decreases with distance with a power index of order unity. The expansion velocity is constant across the torus, at $\sim$1.6 kms$^{-1}$. The polar regions host outflows that are well described by parabolic meridian trajectories joining smoothly between the torus and the star axis with a constant wind velocity of the order of 6 to 7 kms$^{-1}$. A very simple model has been proposed, giving quite a good description of the measured Doppler velocities.

The position angle of the star axis is $\sim$13$^\circ$ east from north and its inclination angle with respect to the sky plane is $\sim$8$^\circ$, both in good agreement with values obtained by studies made of visible and infrared observations.

 Important deviations from a fully symmetric model have been revealed: an excess of the effective density has been observed in both \mbox{CO(3-2)} and \mbox{CO(6-5)} data on the eastern limb of the bicone, particularly strong in the south-east direction. No obvious physics interpretation could be given.
 The continuum observations are consistent with a compact dust source elongated along the equator and with the SED values available in the literature.

 Existing models of the Red Rectangle, such as proposed by Koning et al. (2011) or Men'shchikov et al. (2002), are essentially based on the dust properties and leave much freedom for the gas morphology and kinematics. Yet, the observations presented here are in qualitative agreement with the general picture proposed by Men'shchikov et al. (2002).

 The observations analysed in the present work were done in an early phase of ALMA operation, with only 21 to 24 active antennas in a configuration that extended up to only 380 m. New observations using the complete array and a broader bandwidth would obviously very much improve the quality of the data and their significance for the understanding of the morphology and kinematics of the gas surrounding the Red Rectangle.

\section*{ACKNOWLEDGEMENTS}
 We are indebted and very grateful to the ALMA partnership, who are making their data available to the public after a one year period of exclusive property, an initiative that means invaluable support and encouragement for Vietnamese astrophysics. We particularly acknowledge friendly and patient support from the staff of the ALMA Helpdesk. This paper makes use of the following ALMA data: ADS/JAO.ALMA 2011.0.00223.S. ALMA is a partnership of ESO (representing its member states), NSF (USA) and NINS (Japan), together with NRC (Canada) and NSC and ASIAA (Taiwan), in cooperation with the Republic of Chile. The Joint ALMA Observatory is operated by ESO, AUI/NRAO and NAOJ. We express our deep gratitude to Professors Nguyen Quang Rieu and Thibaut Le Bertre for having introduced us to radio astronomy and to the physics of AGB and post-AGB stars. Financial support is acknowledged from the Institute for Nuclear Science and Technology (VINATOM/MOST), the NAFOSTED funding agency under grant number 103.08-2012.34, the World Laboratory, the Odon Vallet Foundation and the Rencontres du Viet Nam.


\label{lastpage}

\end{document}